\documentclass{iopjournal}
\usepackage{ragged2e}

\newcommand{\mod}[1]{\textcolor{black}{#1}}  

\begin{document}

\articletype{Paper} 

\title{3D modelling of thermal loads during unmitigated vertical displacement events in ITER and JET}

\author{F.J. Artola${^{1,*}}$\orcid{0000-0001-7962-1093}, 
A. Redl${^2}$\orcid{0000-0002-7254-3501},
S.N. Gerasimov${^3}$\orcid{0009-0002-3793-7211},
R.A. Pitts${^1}$\orcid{0000-0001-9455-2698},
I.S. Carvalho${^1}$\orcid{0000-0002-2458-8377},
M. Kong${^4}$\orcid{0000-0002-2004-3513},
G. Simic${^1}$\orcid{0000-0001-9314-9784},
A. Loarte${^1}$\orcid{0000-0001-9592-1117},
J. Van Blarcum${^1}$\orcid{0009-0003-4944-6580},
the \textsc{JOREK} team$^a$, 
the JET contributors$^b$
and the EUROfusion Tokamak Exploitation Team$^c$}

\affil{$^1$ITER Organization, Route de Vinon sur Verdon, 13067 St Paul Lez Durance Cedex, France}

%
\affil{$^2$Max Planck Institute for Plasmaphysics, Boltzmannstr. 2, 85748 Garching, Germany}

\affil{$^3$UKAEA, Culham Campus, Abingdon, Oxon, OX14 3DB, United Kingdom}

\affil{$^4$Ecole Polytechnique Fédérale de Lausanne (EPFL), Swiss Plasma Center (SPC), CH-1015 Lausanne, Switzerland}

\affil{$^*$Author to whom any correspondence should be addressed.}

\affil{$^a$See the author list of M. Hoelzl et al 2024 Nucl. Fusion 64 112016.}

\affil{$^b$See the author list of C.F. Maggi et al 2024 Nucl. Fusion 64 112012.}

\affil{$^c$See the author list of N. Vianello et al 2026 Nucl. Fusion (submitted).}

\email{javier.artola@iter.org}

\keywords{disruptions, ITER, JET, VDE, halo current, heat flux, validation}

\begin{abstract}
Predicting three-dimensional thermal loads during tokamak disruptions is essential for ITER yet remains weakly developed. We present a physics-based workflow that couples MHD simulations of vertical displacement events with field line tracing on a realistic 3D first wall model and a transient wall thermal response. The approach is validated against JET discharges with beryllium main chamber armour, reproducing key global dynamics, non-axisymmetric current features, and the occurrence (or absence) of melting, thereby building confidence in the methodology. We then apply the same workflow to ITER-relevant conditions with tungsten (W) armour, consistent with the new 2024 ITER re-baseline, to assess disruption heat loads and their 3D localization. The resulting analysis demonstrates the resilience of the ITER W first wall against these events and provides predictions for the energy deposition and current flow profiles. Beyond these studies, the workflow enables scenario-by-scenario estimates of disruption-induced thermal loading, allowing to assess the disruption-budget consumption for these events in future devices.
\end{abstract}

\newpage

\justifying

\section{Introduction}

The modelling of 3D thermal loads during tokamak disruptions remains largely unexplored, with no reliable predictive tools for future machines. During the Thermal Quench (TQ) and Current Quench (CQ) phases of disruptions at high plasma currents, the rapid loss of thermal and magnetic energy can cause severe melting of plasma-facing components (PFC) \cite{Lehnen_JNM_2015,Pautasso_NF_1994}, leading potentially to extremely costly operational delays, especially in nuclear devices. Predictive, and as realistic as possible modelling tools, are therefore essential to establish disruption mitigation needs and assess the disruption budget \cite{Lehnen_IAEA_2016} for next step tokamaks. 

\medskip

\begin{figure}[h]
\centering
\includegraphics[width=0.6\textwidth]{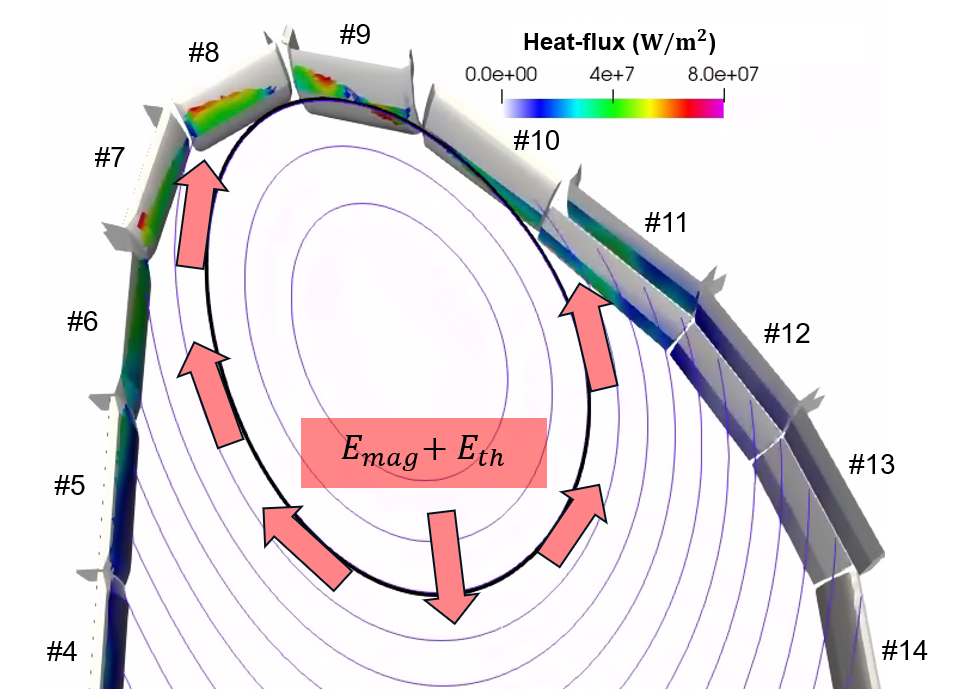} %
\caption{Schematic of the energy flow (red arrows) during an ITER UVDE simulation  (see Sec. \ref{sec:ITER}) together with the first wall panel indexing and the perpendicular heat fluxes on the ITER first wall. The equilibrium is taken at $t=0.391$ s with $q_{95}=1.77$. }
\label{fig:schematic}
\end{figure}

\mod{Among the different types of disruptions, unmitigated vertical displacement events (UVDE) are expected to be the most detrimental to PFCs. In UVDEs, most of the plasma energy is lost through parallel transport along open field lines and the radiated energy fraction is  low. The dominant parallel loss channel focuses thermal loads into narrower regions and yields  higher local energy densities. As a result, PFC damage is much more severe than in radiative disruptions, where the energy is radiated and distributed more uniformly over the PFCs. Fig. ~\ref{fig:schematic} illustrates schematically the energy flow in an ITER UVDE (red arrows).  UVDEs also involve large-amplitude external kink modes, which further increase thermal load localization due to toroidal asymmetries. For a detailed comparison between UVDEs and radiative disruptions, see \cite{Artola_PPCF_2024}.}

\medskip

\mod{
Building on this physical picture, several studies have already quantified ITER CQ  thermal loads for upward-going VDEs \cite{Mitteau_PS_2011,Coburn_NF_2022,Pitts_IAEA_2023}. For the ITER 2016 Baseline, Coburn et al. used the coupled \textsc{DINA--SMITER--MEMOS-U} workflow to evaluate the impact of these events on the beryllium (Be) first wall (FW) armour \cite{Coburn_NF_2022}. That analysis predicted melting for plasma currents above $I_p>7$ MA and severe melt erosion at the nominal 15 MA scenario. In the updated ITER 2024 baseline, Be is replaced by tungsten (W) on the FW \cite{Pitts_NF_2025}, which substantially raises the melt threshold. Even so, axisymmetric \textsc{TOKES} simulations of the same VDE CQ still indicate melting onset around $I_p\approx 10$ MA for W \cite{Pitts_IAEA_2023}.
}

\medskip

\mod{
Despite these advances, existing predictions rely on axisymmetric assumptions that are questionable for the low safety factor ($q_{95}$) conditions typical of such CQs. In \cite{Coburn_NF_2022}, the 3D geometry of the first wall panels (FWP) and the evolving CQ equilibrium were included, but the 2D \textsc{DINA} \cite{Khayrutdinov_JoP_1993} plasma model required a toroidally symmetric parallel heat flux ($q_\parallel$). The \textsc{TOKES} calculations are likewise axisymmetric and, in addition, neglect both the 3D wall structure and the time evolution of the magnetic equilibrium. When $q_{95}$ drops below $\sim 2$, which is typical of these events, large-scale external kink activity takes place. This breaks toroidal symmetry and can generate strong toroidal asymmetries in $q_\parallel$. The consequence is potentially larger heat flux localization than those captured by axisymmetric models, with correspondingly higher peak loads on specific FWPs. It is therefore essential to assess how 3D plasma dynamics modify the localization and magnitude of CQ thermal loads, and to quantify the implications for melting and erosion of the ITER 2024 W FW.
}

\medskip

In this article, we report, for the first time, \textsc{JOREK} simulations of  UVDE heat loads which account for 3D plasma loads using a toroidally shaped  FW geometry (2016 Be ITER baseline). Note the design for the final W FW (DT-1 phase and beyond) has not been completed yet, and , we will assume  that it will retain the main shaping features of the 2016 Be baseline. Using the first-principles methodology described in Section \ref{sec:method}, we simulate two dedicated upward-going UVDEs in JET and validate the predicted thermal and current loads in Section \ref{sec:JET}. This successful validation increases confidence in the model and motivates its application to ITER, where we perform equivalent simulations and analyse the resulting thermal loads (Sec. \ref{sec:ITER}). Finally, Section \ref{sec:conclusions} summarizes the main findings and outlines directions for future work.

\section{Methodology and model assumptions}
\label{sec:method}

The workflow employed to compute disruption-induced thermal loads proceeds in three stages. 
First, nonlinear resistive-MHD simulations of the VDE are performed with the \textsc{JOREK} code (Sec.~\ref{subsec:MHD_sims}), 
yielding time-resolved 3D magnetic fields and plasma source terms such as parallel heat flux and halo current channels. 
Second, a field line tracing post-processor maps these plasma power and current loads onto a high-fidelity 3D model of the FW (Sec.~\ref{subsec:FLT}). Third, for each wall element, a one-dimensional transient heat-conduction problem is solved in the wall-normal direction, using the mapped heat fluxes (Sec.~\ref{subsec:wall_T}). From this, the peak surface temperature is evaluated to assess whether melting occurs.  Note that the wall thermal response is evaluated in post-processing, without feedback on the MHD evolution that may take place once wall erosion arises (due to release of armour material into the quenching plasma).

\subsection{The \textsc{JOREK} MHD simulations}
\label{subsec:MHD_sims}

\mod{With the exception of the new JET case presented in
Section~\ref{sec:JET}, all MHD simulations used in this work were previously reported in \cite{Artola_PPCF_2024}. In the present article, we take the corresponding output data and post-process it to quantify the thermal loads on PFCs. Since the MHD dynamics and the full set of modelling choices are described in \cite{Artola_PPCF_2024}, they are not repeated here in detail. Instead, we briefly recall the model assumptions below and discuss what is their impact on the heat flux evaluation.}

\subsubsection{MHD model set-up}
\mod{The plasma is time-evolved in \textsc{JOREK} \cite{huysmans_NF_2007,Hoelzl_NF_2021} in the reduced-MHD approximation, coupled to the thin-wall code \textsc{STARWALL} \cite{Merkel_arx_2015,hoelzl_JoP_2012} to capture the plasma-vessel electromagnetic interaction via free-boundary conditions for the magnetic field}. The solved variables are $(\psi, \Phi, j, w, T)$,  denoting the poloidal magnetic flux, electrostatic potential, toroidal current density, toroidal vorticity and a single temperature ($T_i=T_e=T/2$). Impurity radiation physics and self-consistent plasma-wall interaction (beyond the plasma-vessel feedback) are not included, and ferromagnetic effects from the JET iron core are omitted. Additionally, the plasma density is constant in time and space, and the parallel main ion flow term is neglected ($v_\parallel=0$). \mod{Under these assumptions, the pre-TQ thermal energy ($E_{th}$) and the magnetic energy converted into ohmic heating during the CQ ($E_{ohm}$), are exhausted through parallel conduction and perpendicular convection.}

\medskip

These modelling choices reproduce global JET observables (e.g. $I_p$, $q_{95}$ and vertical current centroid position ($Z_{curr}$))  within experimental bounds (see \cite{Artola_PPCF_2024} and Section \ref{sec:JET}). For ITER, the same physics model was used but with a rescaling of diffusive times to bring the very long unmitigated VDE evolution into tractable runtimes (re-scaling factor of 60). In addition, all simulations are resolved considering four toroidal Fourier harmonics ($n\in [0,3]$).

\subsubsection{Boundary conditions and their implications for thermal loads}

Dirichlet boundary conditions are used for all variables except for $\psi$ and $j$ where the \textsc{JOREK-STARWALL} coupling is implemented. These simplifications result in the no flow condition ($\mathbf{v}\cdot\mathbf{n}=0$), leading to the following heat flux form at the plasma-wall interface

\begin{equation}
\mathbf{q} \;\simeq\; \mathbf{q}_{\parallel}
= -\,\kappa_{\parallel}\!\big(\overline{T}\big)\,\nabla_{\parallel} T,
\qquad
\overline{T} \equiv \max\!\left(T,\,T_{\min,\kappa_{\parallel}}\right),
\end{equation}
where $\kappa_{\parallel}(T)$ is the Spitzer–Härm conductivity with a floor value defined by $T_{\min,\kappa_{\parallel}}$. \mod{The  Dirichlet condition for the temperature is chosen in all simulations as $T_e=1$ eV.  This choice prevents the far scrape-off layer (SOL) from becoming artificially conducting. A conducting far SOL can spuriously enhance VDE vertical stabilization, effectively mimicking a conducting wall \cite{Krebs_PoP_2020}. In reality, the far SOL is expected to be weakly conducting (vacuum-like)  because of the low plasma density in this area leads to a low ion flux and thus a limitation of the parallel heat flux through the plasma sheath \cite{Artola_PPCF_2021}. Note that this boundary condition is local and does not fix the overall halo resistance. Parallel $T_e$ gradients develop near the \textsc{JOREK} boundary and the halo resistance is set by the internal temperature solution obtained from the energy balance equation. To resolve these boundary-layer gradients numerically without artificially increasing core heat losses, we evaluate $\kappa_{\parallel}$ using a floor value $T_{e,\min,\kappa_{\parallel}}=30~\mathrm{eV}$. This floor is applied only in the computation of $\kappa_{\parallel}$; the temperature field itself is not clipped.}


\medskip

At first sight, these boundary choices may appear too crude for a quantitative heat flux study. However, the JET validation in Section~\ref{sec:JET} demonstrates good agreement for two discharges that span a broad range in toroidal field ($B_T$), $I_p$, and density ($n_e$). This indicates that the CQ conditions place the SOL in a conduction-limited regime, where results are insensitive to the precise wall boundary value of $T_e$. In a conduction-limited SOL, the relevant control parameter is the Knudsen number $\mathrm{Kn}\equiv \lambda_e/L_{\parallel}$, where $\lambda_e$ denotes the electron collision mean-free-path and $L_\parallel$ the field line connection length to the wall. Using representative CQ values from our simulations, $T_{e,\mathrm{core}}=100~\mathrm{eV}$, $n_e=8\times10^{19}\,\mathrm{m^{-3}}$, and $L_{\parallel}=100~\mathrm{m}$, we find $\lambda_e\simeq 1.4~\mathrm{m}$ and hence $\mathrm{Kn} \approx 0.014\ll 0.1$. This is consistent with steep parallel temperature gradients set by Spitzer–Härm conduction and weak sensitivity to the boundary condition \cite{stangeby2000plasma}. A fully self-consistent set of boundary conditions, such as those described in \cite{Artola_PPCF_2021}, remains numerically intractable in \textsc{JOREK} for the present type of disruption simulations \cite{Artola_PPCF_2024}. It is important to emphasize that partial implementation of such models is insufficient. For example, imposing sheath boundary conditions on the temperature alone still leaves the density at the plasma–wall interface as a free parameter, which can easily take higher values than pre-disruptive levels \cite{Artola_PPCF_2021}. If the density is instead evolved self-consistently, including effects such as neutral recycling, then sheath boundary conditions on the electric potential must also be imposed in order to prevent the formation of spurious currents in regions with low ion density, which limits the ion flux and current flow as discussed above.

\subsection{Field line tracing}
\label{subsec:FLT}
\begin{figure}[htbp]
\centering
\includegraphics[width=0.98\textwidth]{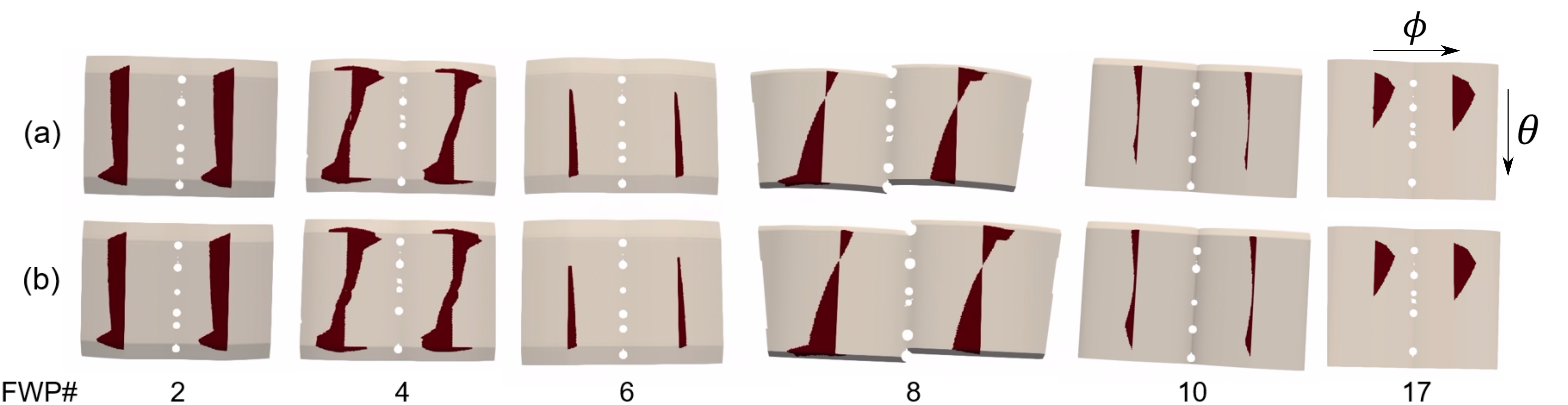} 
\caption{Wetted area (red) calculated in \textsc{SMITER} (a) and \textsc{JOREK} (b) for 6 out of the 18 poloidally distributed FWPs for a pre-disruptive ITER X-point plasma (IMAS URI = imas:mdsplus?user=public;pulse=135011;run=7;database=ITER;version=3 at 200s).}
\label{fig:SMITER_bench}
\end{figure}

Once $q_{\parallel}$ is obtained from the MHD simulations, it is mapped from the axisymmetric
\textsc{JOREK} boundary to the 3D wall elements. The incident heat flux, $q_{\perp}$, is then computed by projecting the parallel heat flux onto the local wall-normal vector, $q_{\perp} = \mathbf{q}_{\parallel}\!\cdot\!\mathbf{n}$, but only for wall elements identified as wetted by the plasma. The wetted elements are determined via field line tracing as performed in \cite{Coburn_NF_2022}, following the same procedure as in the \textsc{SMITER} code \cite{Kos_FED_2019}: a wall element is considered wetted if its field line connects to the plasma without intersecting another wall element within a predefined connection length. For the simulations employed in this work, the predefined minimum connection length for plasma wetting is 5 m. As a consistency check, \textsc{JOREK} was successfully benchmarked against \textsc{SMITER} for the wetted-area calculation on the same 3D CAD model of the ITER FWPs as depicted in Fig. \ref{fig:SMITER_bench}.

\subsection{Wall temperature evolution}
\label{subsec:wall_T}

For each wall element, the transient temperature evolution is obtained by solving an independent one-dimensional heat-diffusion equation along the local surface normal. This treatment is intentionally conservative: neglecting lateral (toroidal or poloidal) heat spreading tends to slightly overestimate the peak surface temperature. The thermal model employs temperature-dependent specific heat capacity and thermal conductivity for both W (ITER) and Be (JET), while the weak temperature dependence of the solid density is neglected. At the plasma-facing surface, the boundary condition prescribes the heat flux
density $q_{\perp}$ computed in Section ~\ref{subsec:FLT}:
\[
\frac{\partial T}{\partial x}\bigg|_{\mathrm{front}}
= -\,\frac{q_{\perp}(t)}{k\!\left(T\right)},
\]
where $x$ is the coordinate normal to the surface and $k(T)$ is the temperature-dependent thermal conductivity.
At the rear surface, a zero-flux (thermally insulated) boundary condition $\partial T/\partial x=0$ is applied. For all simulations, a wall thickness of $12~\mathrm{mm}$ is assumed, consistent with the ITER W FW armour thickness in regions expected to receive the highest disruption loads \cite{Pitts_NF_2025}. Although the Be tiles in JET are thicker, neither this difference nor the rear-side insulation assumption significantly affects the results, since the disruption duration is much shorter than the characteristic thermal diffusion time through the material. As a consistency check, the numerical solver was verified against the analytical solution for a fast, constant heat pulse in the semi-infinite slab approximation, yielding excellent agreement.

\section{Model validation in JET VDEs}
\label{sec:JET}
In this section we apply the workflow detailed in Section \ref{sec:method} to two different JET upward-going UVDEs, namely pulse numbers \#95110 and \#84832 described in \cite{Gerasimov_PS_2024}, with pre-disruptive $I_p$, $B_T$, and $n_e$ of ($1.1$ MA, $1.2$ T, $1.5\times 10 ^{19}$ m$^{-3}$) and ($2.2$ MA, $2.2$ T, $8.0\times 10 ^{19}$ m$^{-3}$) respectively.

\begin{figure}[htbp]
\centering
\includegraphics[width=0.95\textwidth]{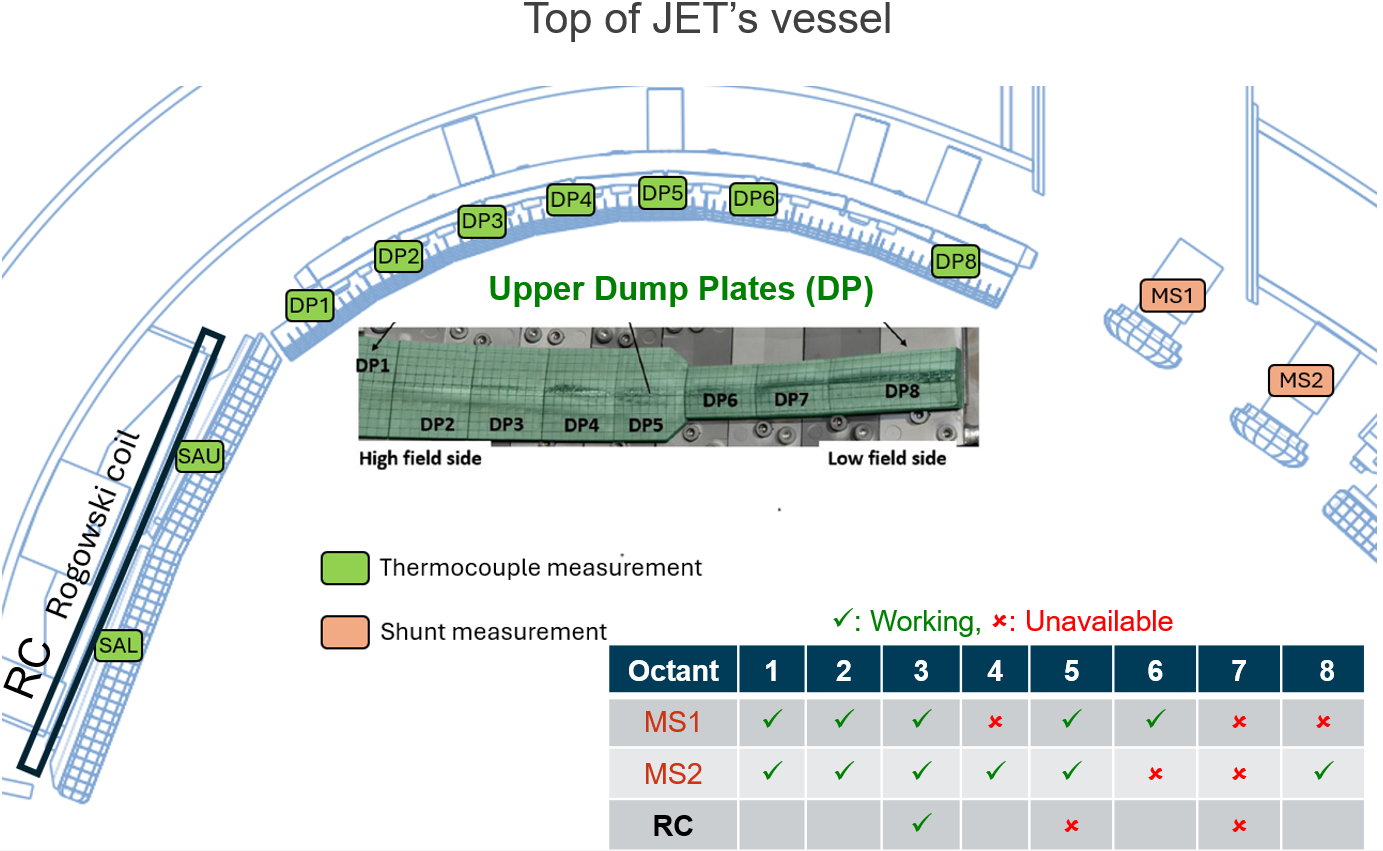} 
\caption{Summary of the JET diagnostics used for thermal loads and halo currents in this study. Green boxes denote TC measurements while light-red boxes denote shunt measurements (MS1/2). The black and thick rectangular outline shows a Rogowski coil (RC) and the bottom table shows the toroidal coverage and availability for halo current measurements in each of the JET octants for the two pulses analysed here. }
\label{fig:JET_diagno}
\end{figure}

\begin{figure}[h]
\centering
\includegraphics[width=0.4\textwidth]{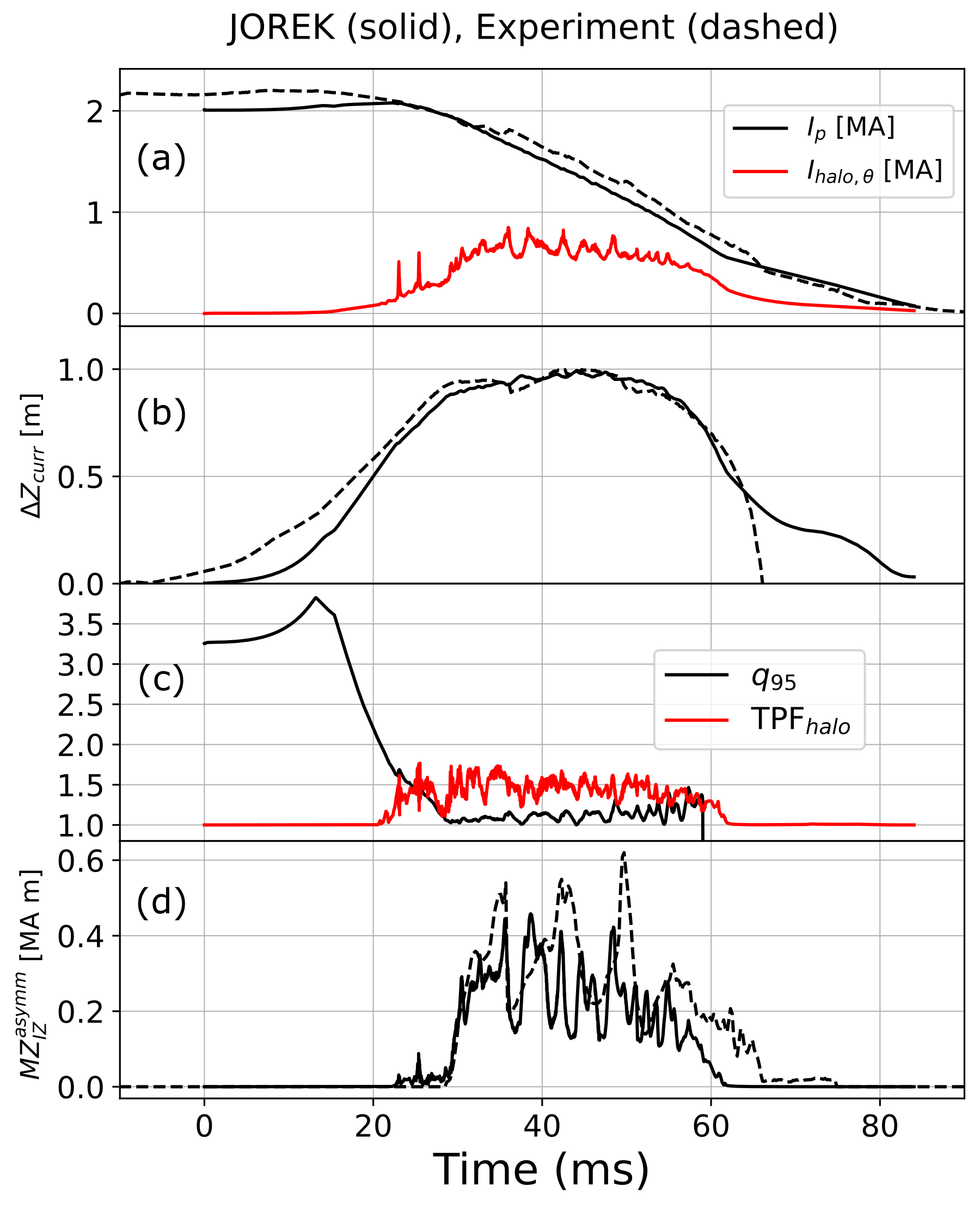} 
\includegraphics[width=0.56\textwidth]{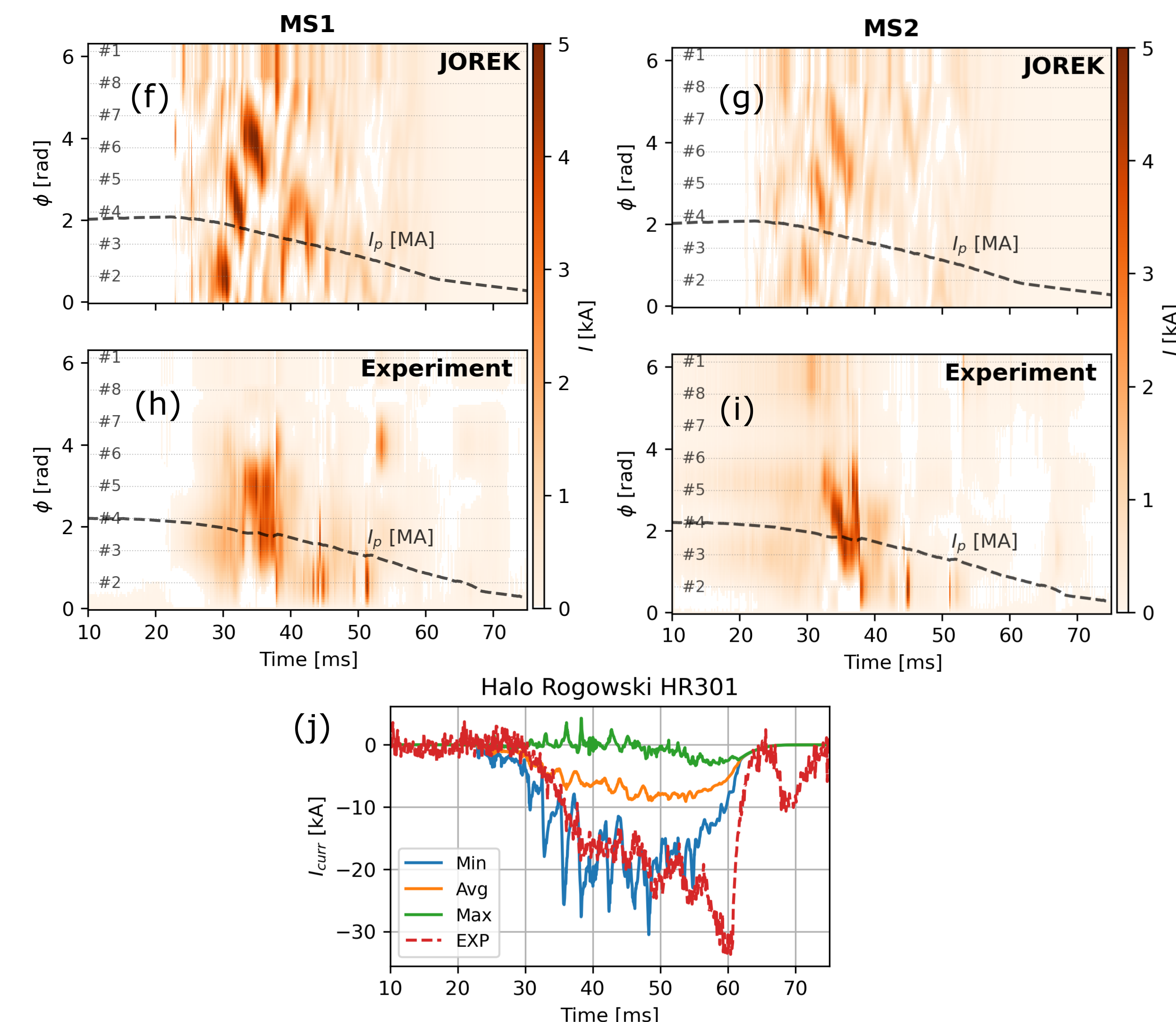}
\caption{(Left) Comparison of simulated values (solid) with experimental values (dashed) for pulse \#84832. (a) Plasma current and poloidal halo currents. (b) Vertical displacement of the current-centroid. (c) Edge safety factor and toroidal peaking factor of the poloidal halo currents. (d) $n=1$ asymmetry of the vertical current moment as defined in \cite{Gerasimov_NF_2014}. (Right) Toroidal current distribution measured by the mushroom shunts MS1 (f,h) and MS1 (g,i) in \textsc{JOREK} (f,g) and in the experiment (h,i). The indices (e.g. \# 8) show the octant on which the measurement is taken. Panel (j) shows the halo current measured by Rogowski coil HR301 (dashed red) and the minimum, average, and maximum currents obtained from a set of equivalent toroidally distributed coils in \textsc{JOREK}. Note that the time has been shifted and that $t=0$ corresponds to 51.5517s in the JET shot.} 
\label{fig:JET_ov}
\end{figure}

\subsection{Measurement diagnostics in JET}
\label{subsec:measurements}
Measurements of $I_p$, $Z_{curr}$, and their toroidal asymmetries are obtained using the magnetic diagnostics and analysis methods described in \cite{Gerasimov_NF_2014}, which were also employed for comparison with \textsc{JOREK} simulations in \cite{Artola_PPCF_2024}. For the thermal load analysis, sub-surface thermocouples (TC) are located in the Be upper Dump Plates (DP) and in the Be upper inner-wall protection limiters (``sausage'' limiters, SAU/SAL). The upper DP TCs are located in Octant~2, whereas the sausage-limiter TCs are located in Octant~1. The poloidal positions of the TC can be seen in Fig \ref{fig:JET_diagno}.  Because UVDEs are toroidally asymmetric \cite{Gerasimov_NF_2014},  measurements from a single toroidal sector can be misleading. To assess the degree of toroidal asymmetry, we also  use  halo current data from the so-called ``mushroom'' shunt (MS) diagnostics, also shown in Fig.~\ref{fig:JET_diagno}. These shunts are installed in several octants, although some channels are unavailable for the pulses modelled here due to technical issues. Since halo currents and parallel heat fluxes to the wall are both driven by the same MHD and connection length geometry, a correlation between the two is expected. The shunt signals therefore provide a proxy for the toroidal distribution (and effective width) of the heat deposition in regions not instrumented with TCs.

\medskip

The TCs are used to reconstruct the total energy deposited on the  PFCs using the calorimetric inversion procedure described in \cite{Matthews_NME_2017}. Although the upper DPs are instrumented with sub-surface TCs, the calorimetric analysis requires internal thermal equilibrium within the tile volume, using long tile cooling times for fitting. As a result, the inferred energy deposition cannot be separated into pre-disruptive and disruption deposited energy with this method, nor can the energy fluxes during the TQ and CQ be distinguished. To isolate the disruption contribution, we subtract the energy measured in a reference discharge that is either non-disruptive or strongly mitigated. For discharge \#95110, the reference is \#95108 \cite{Gerasimov_PS_2024}, in which a neon shattered pellet was injected prior to the VDE. For \#84832, the reference is the non-disruptive pulse with very similar characteristics (\#85364).

\subsection{MHD simulation for discharge \#84832}
JET discharge \#84832 has been simulated with \textsc{JOREK} using the same physics model and numerical parameters as for discharge
\#95110 \cite{Artola_PPCF_2024}. Experimentally, this pulse underwent an H-L back-transition followed by a loss of vertical stability \cite{Gerasimov_PS_2024}. In contrast to \#95110, discharge \#84832 produced melting of the upper DPs and visible release of Be droplets \cite{Jepu_NF_2019}. This can be attributed to the substantially higher stored energies: the magnetic energy was about four times larger, and the pre-TQ thermal energy was $2.2$~MJ compared to only $0.21$~MJ in \#95110. These features make \#84832 a particularly relevant case for assessing thermal loads.

\medskip

Figs. ~\ref{fig:JET_ov} (a–d) compare key global quantities between simulation and experiment:  $I_p$, $Z_{curr}$ and the $n=1$ asymmetry of the vertical current moment (as defined in Ref.~\cite{Gerasimov_NF_2014}). In all cases, the simulated evolution follows the experimental trends closely, giving confidence in the modelled global dynamics. The toroidal distribution of halo currents is shown in Fig.~\ref{fig:JET_ov} (f–i), where the mushroom shunt measurements are compared with the field line tracing calculation from \textsc{JOREK}. The simulations capture both the sign and magnitude of the measured currents, although the toroidal pattern in \textsc{JOREK} exhibits  additional rotation  (one full turn instead of a half turn). The direction of this rotation differs from the experiment and will require further investigation. Finally, Fig.~\ref{fig:JET_ov} (j) compares the halo current measured by the Rogowski coil HR301 with the range of values obtained from a set of virtual toroidally distributed coils in \textsc{JOREK}. The simulated signals reproduce the experimental magnitude when the toroidal sector corresponding to the maximum current (in absolute magnitude) is considered.

\medskip

\mod{Regarding the duration of the energy deposition in the simulations, the TQ lasts about $0.54$ and $2.7~\mathrm{ms}$ for \#95110 and \#84832, respectively. We define the TQ duration from the thermal energy decay as $ \tau_{\mathrm{TQ}}^{20\textrm{--}90} = (t_{0.2}-t_{0.9})/0.7$
where $t_X$ denotes the time at which $E_{th}(t)=X\,E_{th}(t=0)$. A reliable estimate of the experimental TQ duration was not possible for these discharges, since the TQ is not clearly resolved in the available signals. For the CQ, using the standard $\tau_{\mathrm{CQ}}^{20\textrm{--}80}$ definition, we obtain $16$ and $52~\mathrm{ms}$ for \#95110 and \#84832, respectively.}

\begin{figure}[h]
\centering
\includegraphics[width=0.49\textwidth]{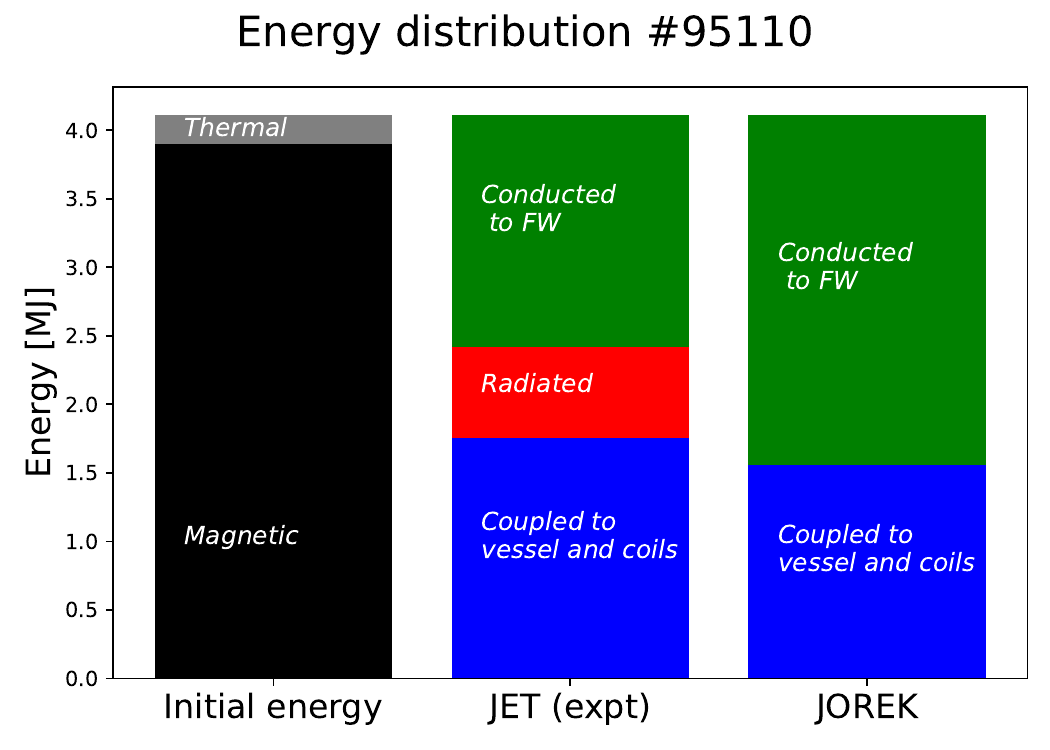} %
\includegraphics[width=0.49\textwidth]{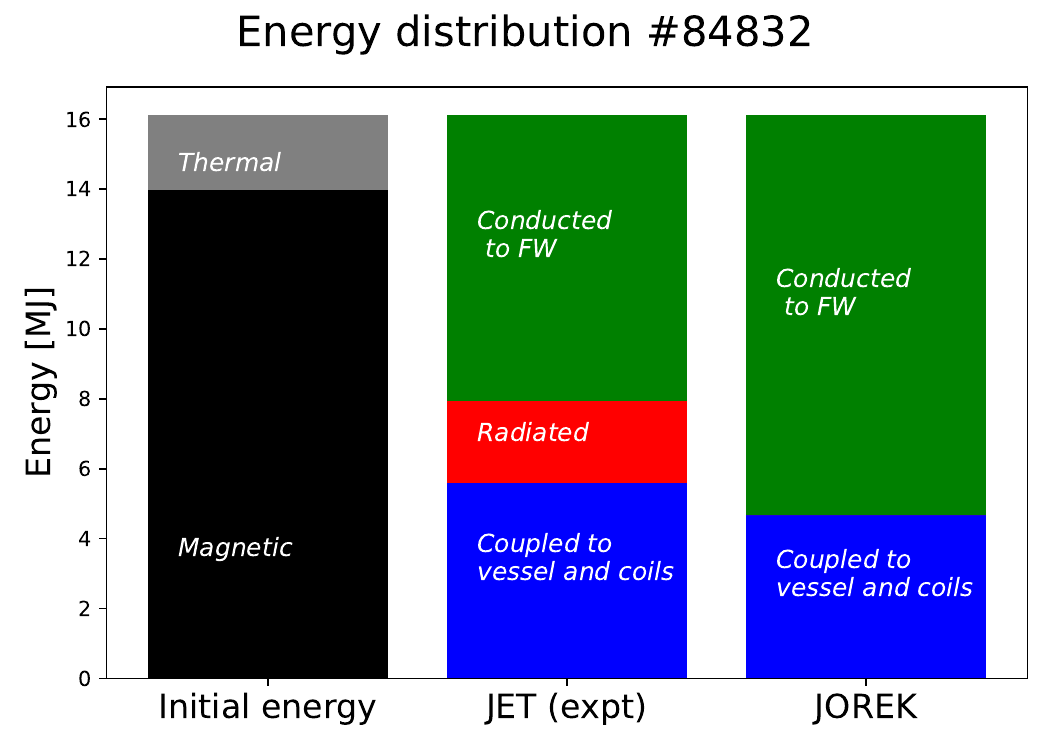} 
\caption{\mod{Global energy balance for \#95110 (left) and \#84832
(right). The first column shows the pre-disruptive stored energy, split
into thermal energy (grey) and poloidal magnetic energy (black,
$0.5\,L\,I_p^2$). The second column (JET) shows the experimental energy
partition: radiated energy from KB5V (red), energy coupled to external
conductors estimated from \cite{Lehnen_NF_2013} using the CQ duration
(blue), and the residual conducted energy to PFCs (green). The third
column (\textsc{JOREK}) shows the corresponding partition from the simulations,
with the coupled contribution computed by integrating the Poynting flux
to external conductors.}}
\label{fig:global_energy_bal}
\end{figure}

\begin{figure}[h!]
\centering
\includegraphics[width=0.49\textwidth]{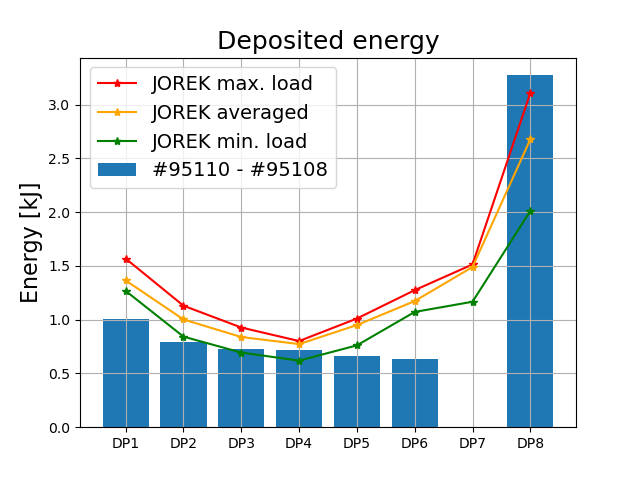} %
\includegraphics[width=0.49\textwidth]{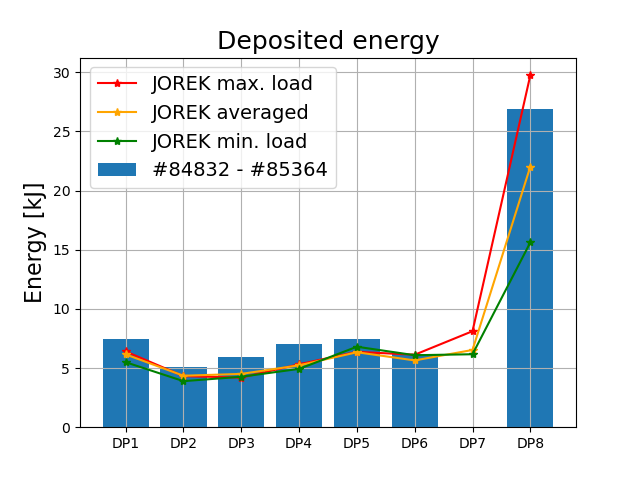} 
\caption{Energy deposited on the upper DPs as measured by the JET TCs using the reference pulse subtraction method explained in Section \ref{subsec:measurements} (blue bars) and that calculated in the \textsc{JOREK} simulations (colored lines) for \#95110 (left) and \#84832 (right). The different simulation lines correspond to the toroidal locations with minimum, average, and maximum deposited energy in the totality of the upper DPs obtained in \textsc{JOREK}. Note that the \textsc{JOREK} results have been scaled down by a factor $(1-f_{rad})$ to account for the radiation losses in the experiment which are not modeled by \textsc{JOREK}. The error in the estimations of the deposited energy is expected to be $<$20\% \cite{Matthews_NME_2017}. }
\label{fig:deposited_energy}
\end{figure}

\subsection{Thermal load analysis}

\mod{The total radiated fraction ($f_{rad}$) during the TQ and CQ, inferred from the vertical bolometer (KB5V) system \cite{Ingesson_JET_2020} is $f_{\mathrm{rad}}=28\%$ for \#95110 and $22\%$ for \#84832. To calculate this fraction, we have evaluated the global energy balance for these discharges following the procedure of \cite{Lehnen_NF_2013}. Fig. ~\ref{fig:global_energy_bal} summarizes the result. The first column shows the pre-disruptive stored energies, namely the thermal energy and the poloidal magnetic energy. The second column shows the experimental partition of this energy into radiation, coupling to external conductors, and conducted energy to PFCs. The coupled fraction (blue) is estimated using the dependence in \cite{Lehnen_NF_2013} as a function of the CQ duration, yielding 45\% for \#95110 and 40\% for \#84832. The conducted energy to PFCs (green) is then obtained as the residual after subtracting the radiated and coupled contributions from the total pre-disruptive energy. The third column shows the corresponding partition from \textsc{JOREK}. In the simulations, the coupled energy is obtained by time-integrating the Poynting flux at the \textsc{JOREK} boundary, while the conducted contribution is evaluated by integrating directly the boundary heat fluxes.}
 
\medskip

Following the workflow described in Section~\ref{sec:method} and the experimental set-up in Section~\ref{subsec:measurements}, the deposited energies in the upper DPs for both discharges are shown in Fig.~\ref{fig:deposited_energy}. Since radiation cooling is not included in the \textsc{JOREK} simulations, the computed heat fluxes have been rescaled by a factor $(1-f_{\mathrm{rad}})$ for a consistent comparison. The different \textsc{JOREK} lines in Fig.~\ref{fig:deposited_energy} correspond to toroidal locations where the total energy deposited on the upper DPs reaches its maximum, minimum and its average value. The results indicate that toroidal asymmetries in the deposited energy are on the order of $\sim20\%$. The largest asymmetry is found in upper DP 8, caused primarily by the TQ phase, which concentrates most of its thermal energy deposition on that plate.

\begin{figure}[h]
\centering
\includegraphics[width=0.49\textwidth]{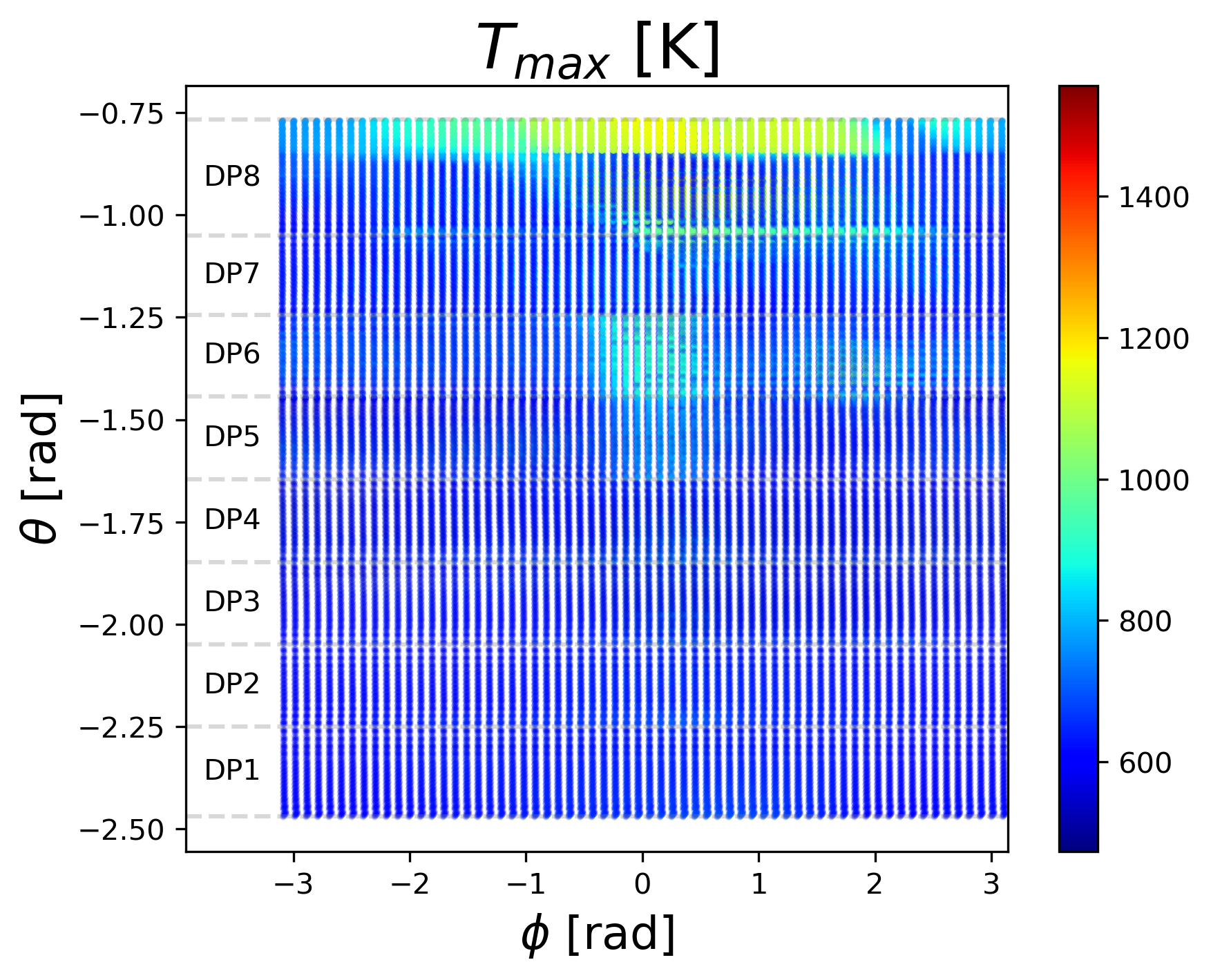} %
\includegraphics[width=0.49\textwidth]{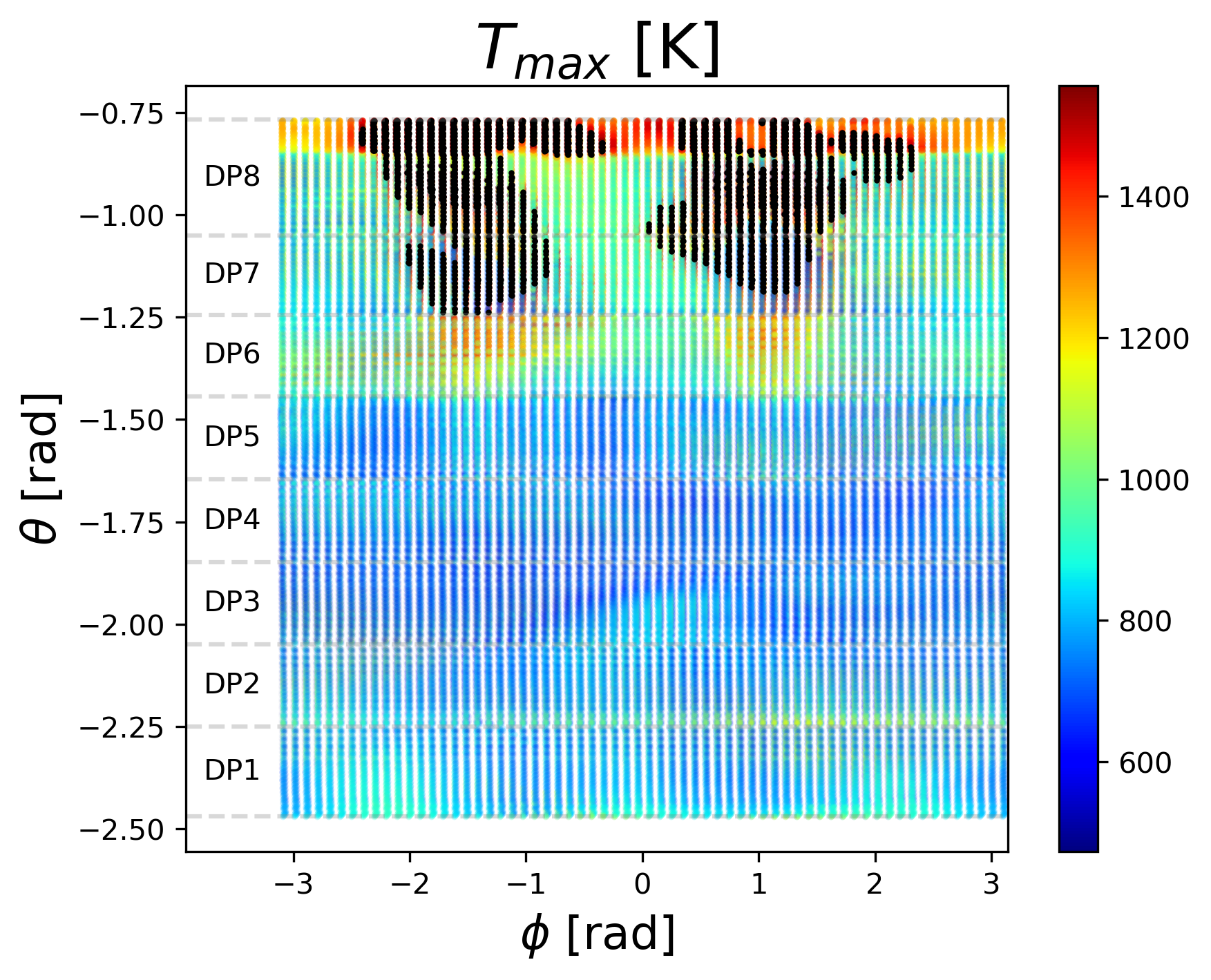} 
\caption{Maximum surface temperature on the JET upper DPs as predicted by \textsc{JOREK} for \#95110 (left) and \#84832 (right). The plates are displayed in toroidal ($\phi$) and poloidal ($\theta$) angles and the black dots represent elements with temperatures beyond the melting point of Be (1556 K). }
\label{fig:Tmax_JET}
\end{figure}

\medskip

Starting with a pre-disruptive wall temperature of $200^{\circ}$C for this discharge \cite{Jepu_NF_2019}, Fig. \ref{fig:Tmax_JET} shows the maximum surface temperature reached in the JET upper DPs as predicted by the \textsc{JOREK} heat flux calculations. For discharge \#95110 (left), no melting is predicted, consistent with the absence of visible melt damage in the experiment. In contrast, for discharge \#84832 (right), melting of Be is predicted on upper DPs~7 and~8 (black points), in agreement with the experimental observation of molten Be release in this region~\cite{Jepu_NF_2019}. The longest predicted melting duration occurs on upper DP~8, where the surface remains above the Be melt temperature ($1556~\mathrm{K}$) for approximately $6$~ms. 
\medskip

\begin{figure}[h]
\centering
\includegraphics[width=0.49\textwidth]{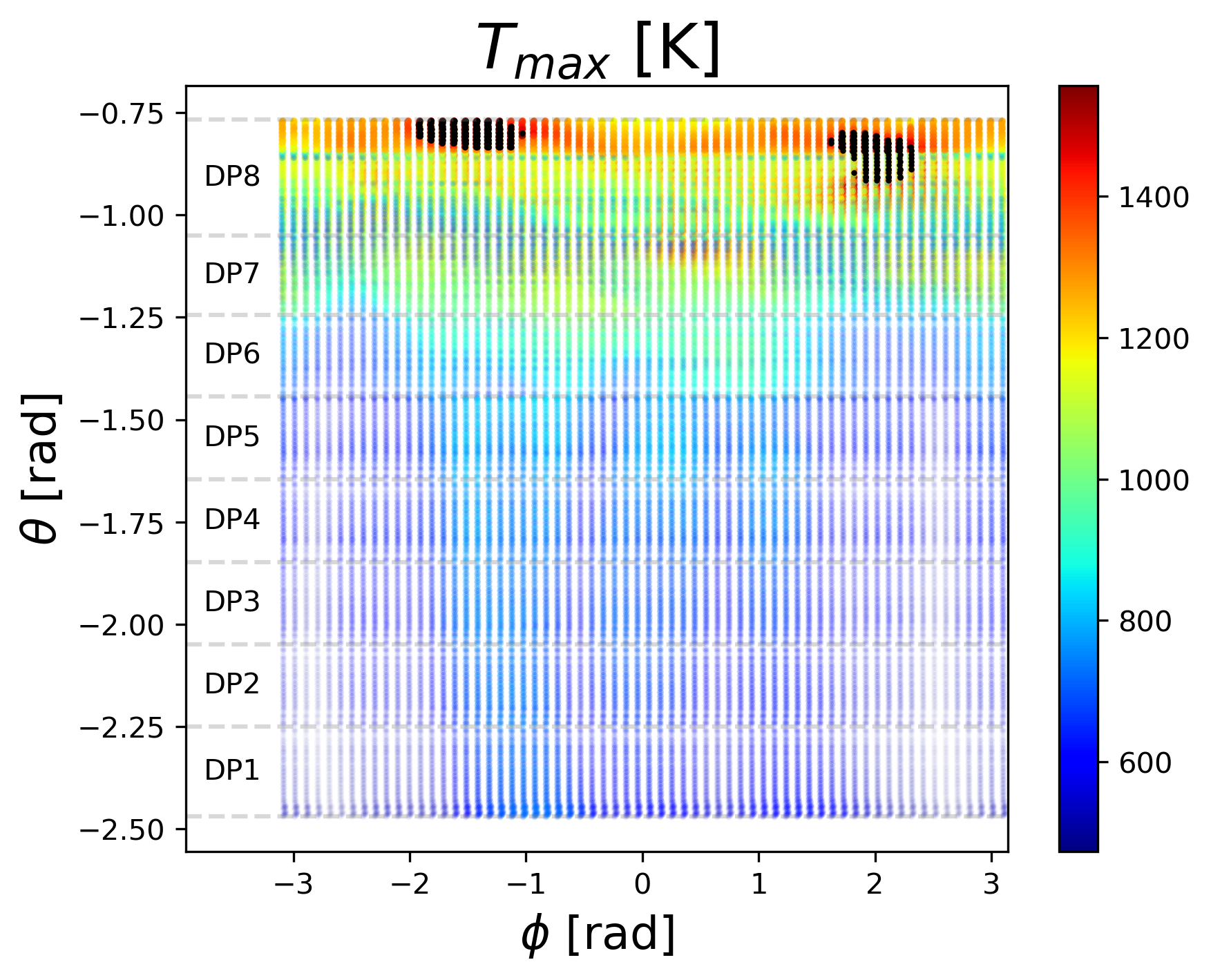} %
\includegraphics[width=0.49\textwidth]{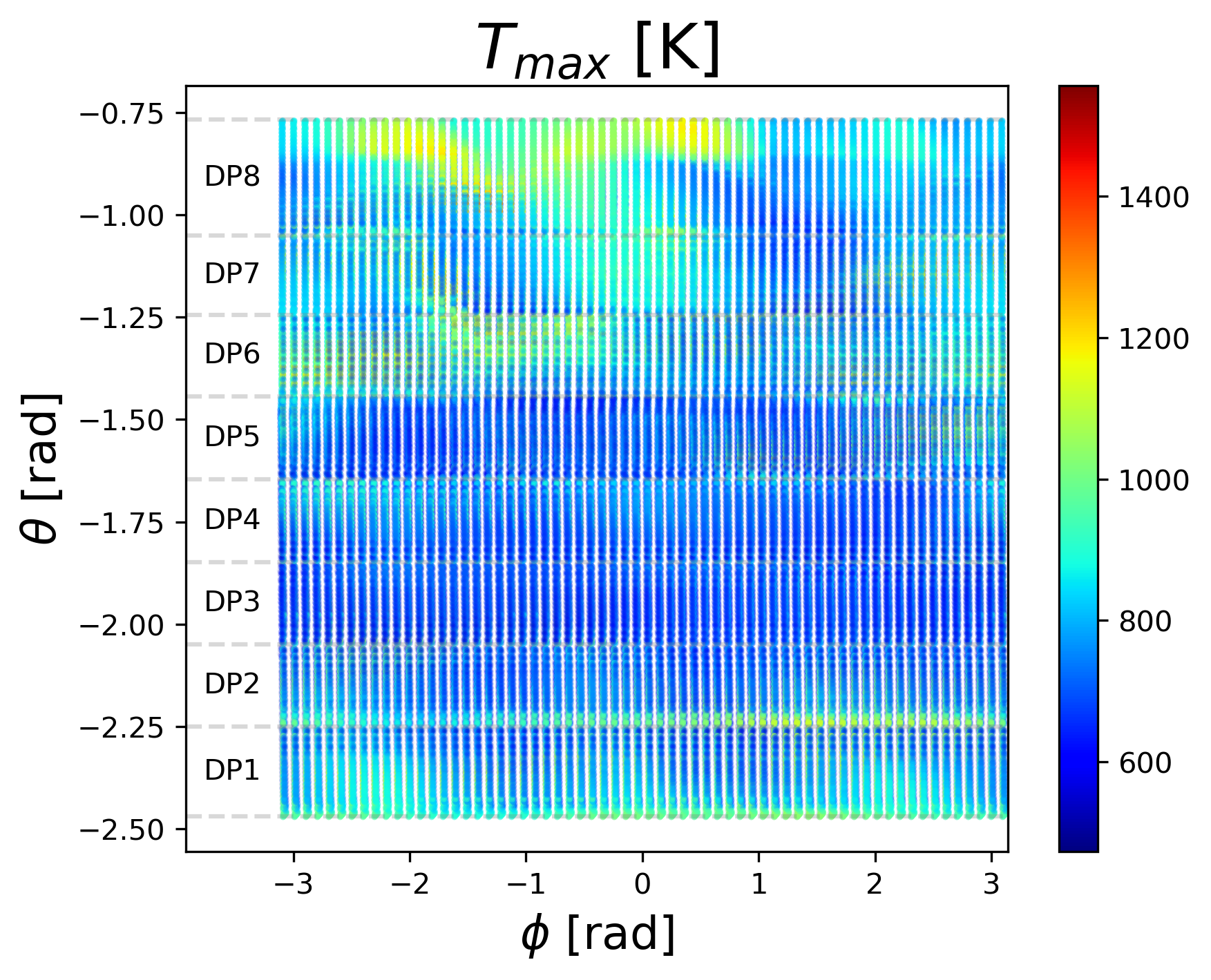} 
\caption{\mod{Maximum surface temperature on the JET upper DPs for pulse \#84832 as predicted by \textsc{JOREK} considering only the TQ (left) and only the CQ (right) . The plates are displayed in toroidal ($\phi$) and poloidal ($\theta$) angles and the black dots represent elements with temperatures beyond the melting point of Be (1556 K).} }
\label{fig:Tmax_JET_sep}
\end{figure}

A more detailed analysis of the \textsc{JOREK} output shows that, for \#84832, neither the TQ nor the CQ phase separately is sufficient to produce significant melting  (recall that the contributions cannot be separated experimentally). \mod{This can be seen in Fig. \ref{fig:Tmax_JET_sep}, where the thermal load analysis is performed considering the TQ and CQ separately}. Considering only the TQ fluxes yields at most mild surface melting, with a duration below $0.6$~ms, while the CQ loads alone do not raise the surface temperature above the melt point. Melting is obtained \textit{only} when both phases are combined: the TQ acts to pre-heat the PFC surface, and the subsequent CQ heat loads then sustain and further increase the surface temperature. As also discussed in \cite{Ratynskaia_NF_2020}, this illustrates the importance of capturing the combined action of both disruption phases for accurate melt predictions in these JET cases. 

\subsection{\mod{Characterization of the deposition area}}
\label{sec:JET_area}
\begin{figure}[h]
\centering
\includegraphics[width=0.41\textwidth]{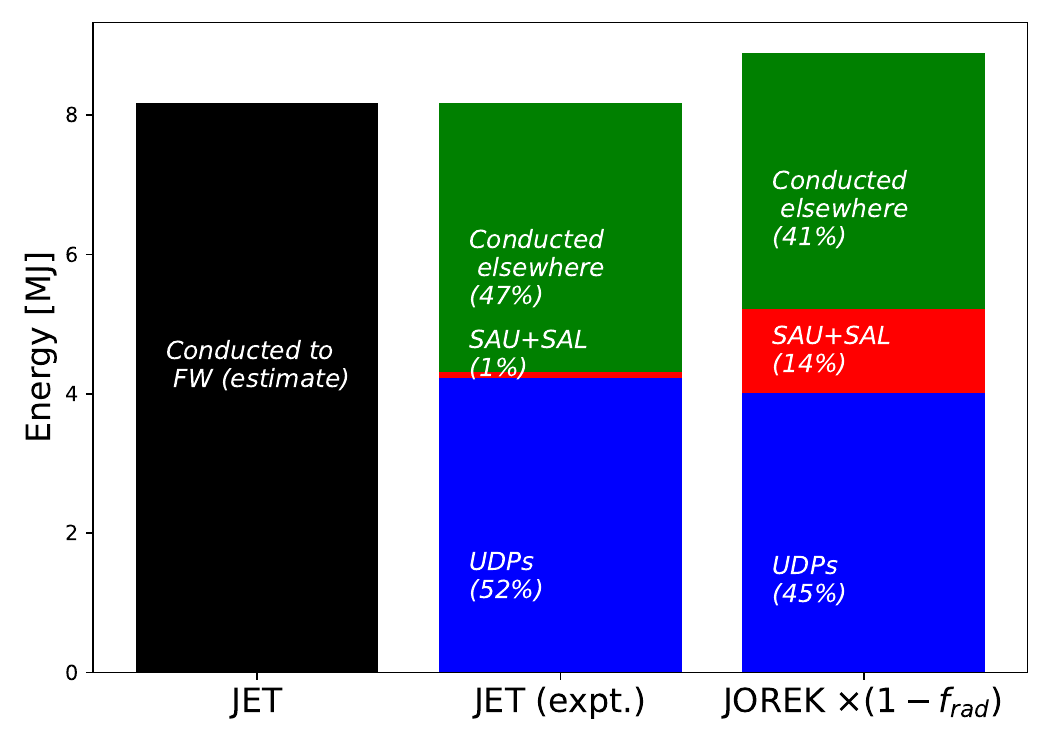} %
\includegraphics[width=0.5\textwidth]{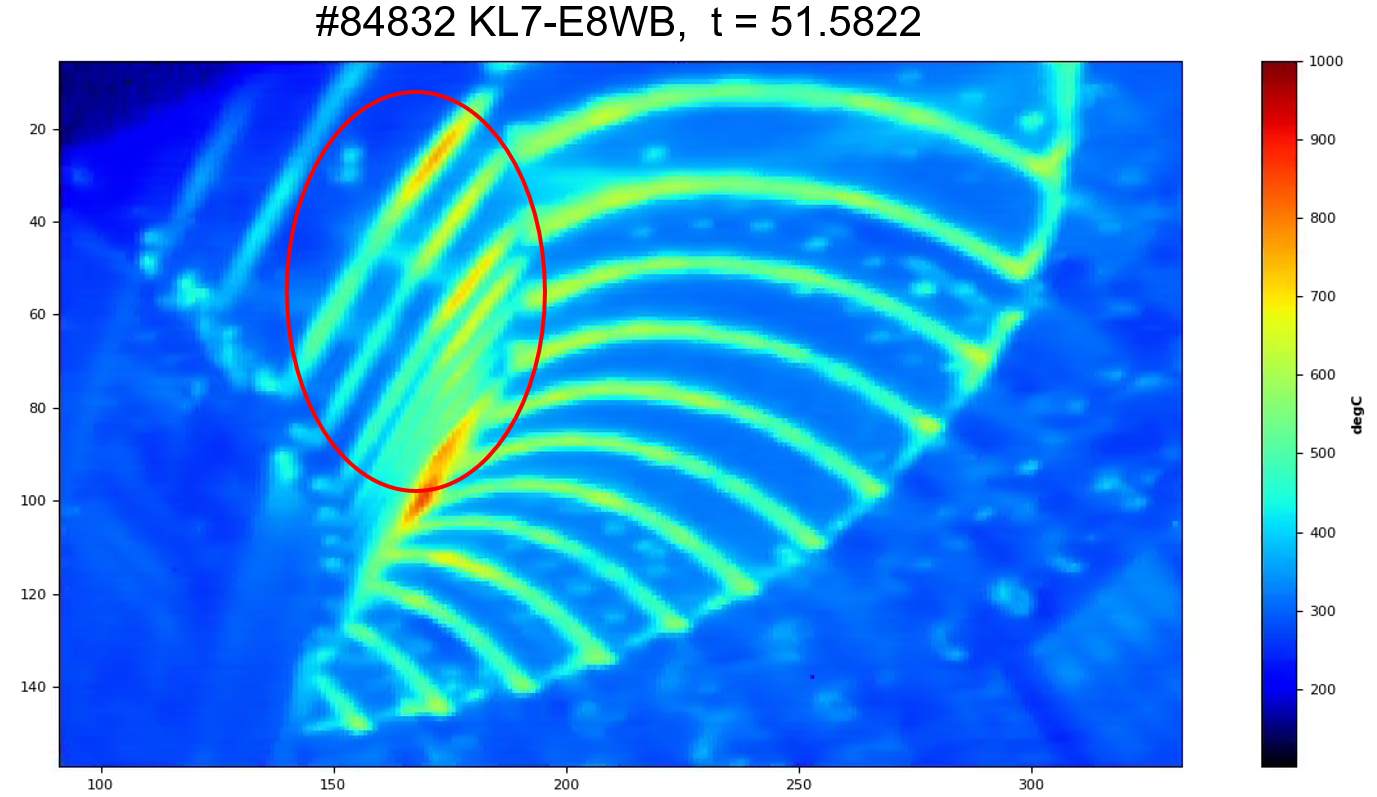} 
\caption{\mod{Energy deposition outside the upper DPs for \#84832. Left: partition of the conducted energy to PFCs, using the total conducted energy inferred from the global energy balance (Fig.~\ref{fig:global_energy_bal}) and the fractions measured by thermocouples at the upper DPs and at the inner-wall protection limiters (SAU/SAL). Right: wide-angle IR image (KL7-E8W8) at $t=51.5822$s ($I_p=87\%I_{p,0}$ toroidally asymmetric phase \cite{Gerasimov_PS_2024}) showing a pronounced heat-load footprint at the SAU/SAL toroidal location (octant~1, marked with a red ellipse), consistent with significant energy deposition in this region.} }
\label{fig:conducted_energy_balance}
\end{figure}

\mod{
A key result of this study is that the conducted energy is deposited over an area that extends well beyond the upper DPs. This is illustrated in Fig.~\ref{fig:conducted_energy_balance} (left), where we compare the total conducted energy inferred from the global energy balance (Fig.~\ref{fig:global_energy_bal}) with the fraction measured by TCs (assuming toroidal symmetry). In particular, we separate the energy measured at the upper DPs from that measured at the inner-wall protection limiters (SAU/SAL). Both the experimental reconstruction and the \textsc{JOREK} simulations indicate that only about half of the conducted energy is deposited on the upper DPs and SAU/SAL locations. The remaining fraction must therefore be deposited on other PFCs, which confirms that the heat-load footprint is broad rather than confined to the upper DP region.}

\medskip

\mod{
A discrepancy appears for the SAU/SAL contribution. The TCs indicate that the energy reaching the SAU/SAL limiters is  $\sim$1\% of the total conducted energy, whereas \textsc{JOREK} predicts a much larger fraction, around 14\%. We suspect that this difference is given by a diagnostic or analysis issue for the SAU/SAL TC signal. This interpretation is supported by the wide-angle IR measurements, which show significant heating in the same region and at the same toroidal location (octant~1) where the SAU/SAL TCs are installed (see red circle in Fig.~\ref{fig:conducted_energy_balance}, right). Moreover, the Rogowski coil collecting current from the SAL/SAU limiters (Fig.~\ref{fig:JET_diagno}, RC) measures substantial current flow. Although it is located in octant~3 (rather than octant~1), its signal is comparable to the simulated value (Fig.~\ref{fig:JET_ov} j). This further supports the interpretation that the SAL/SAU limiters carry appreciable halo current and are therefore expected to experience non-negligible thermal loads.
}

\begin{figure}[h]
\centering
\includegraphics[width=0.49\textwidth]{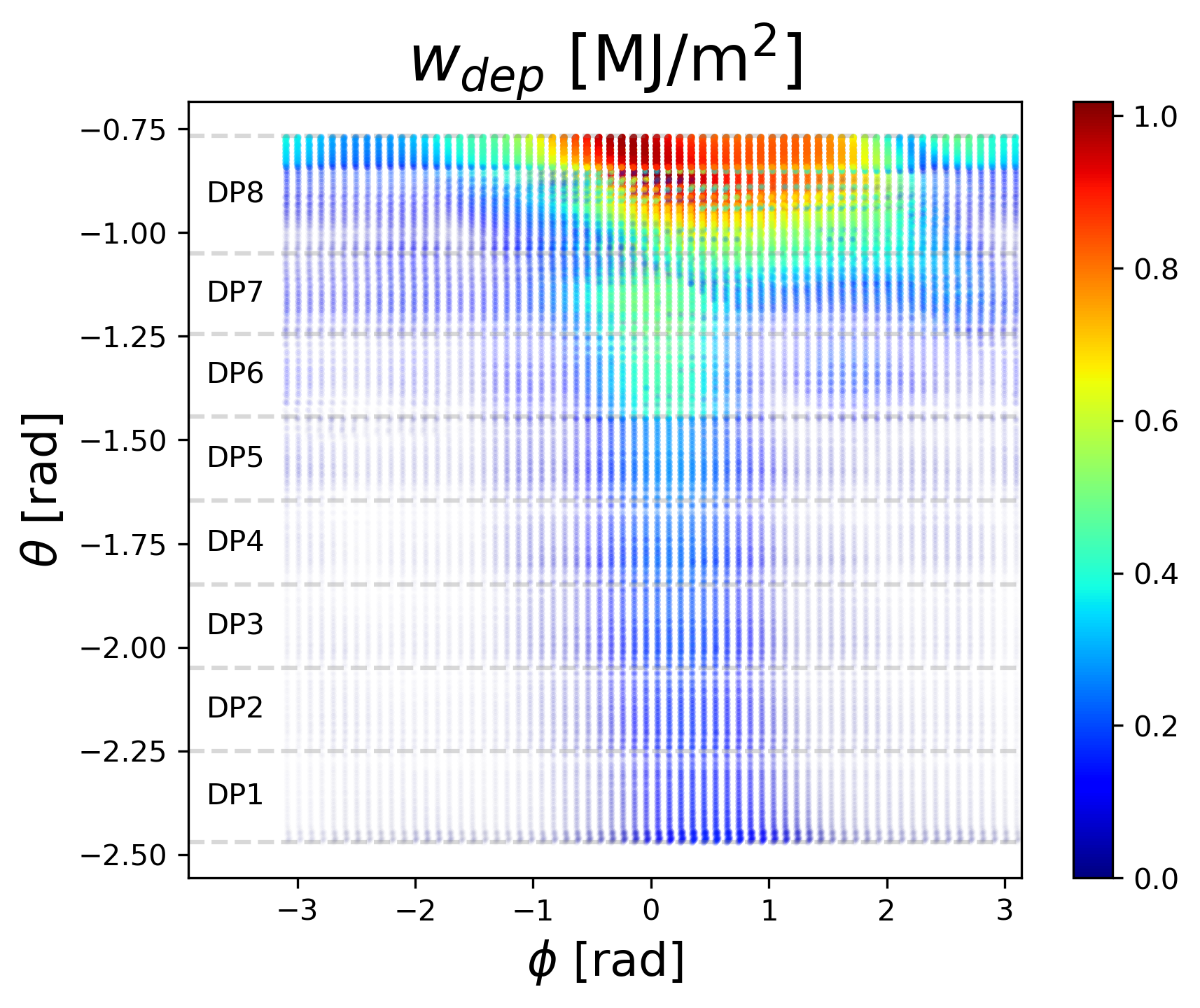} %
\includegraphics[width=0.49\textwidth]{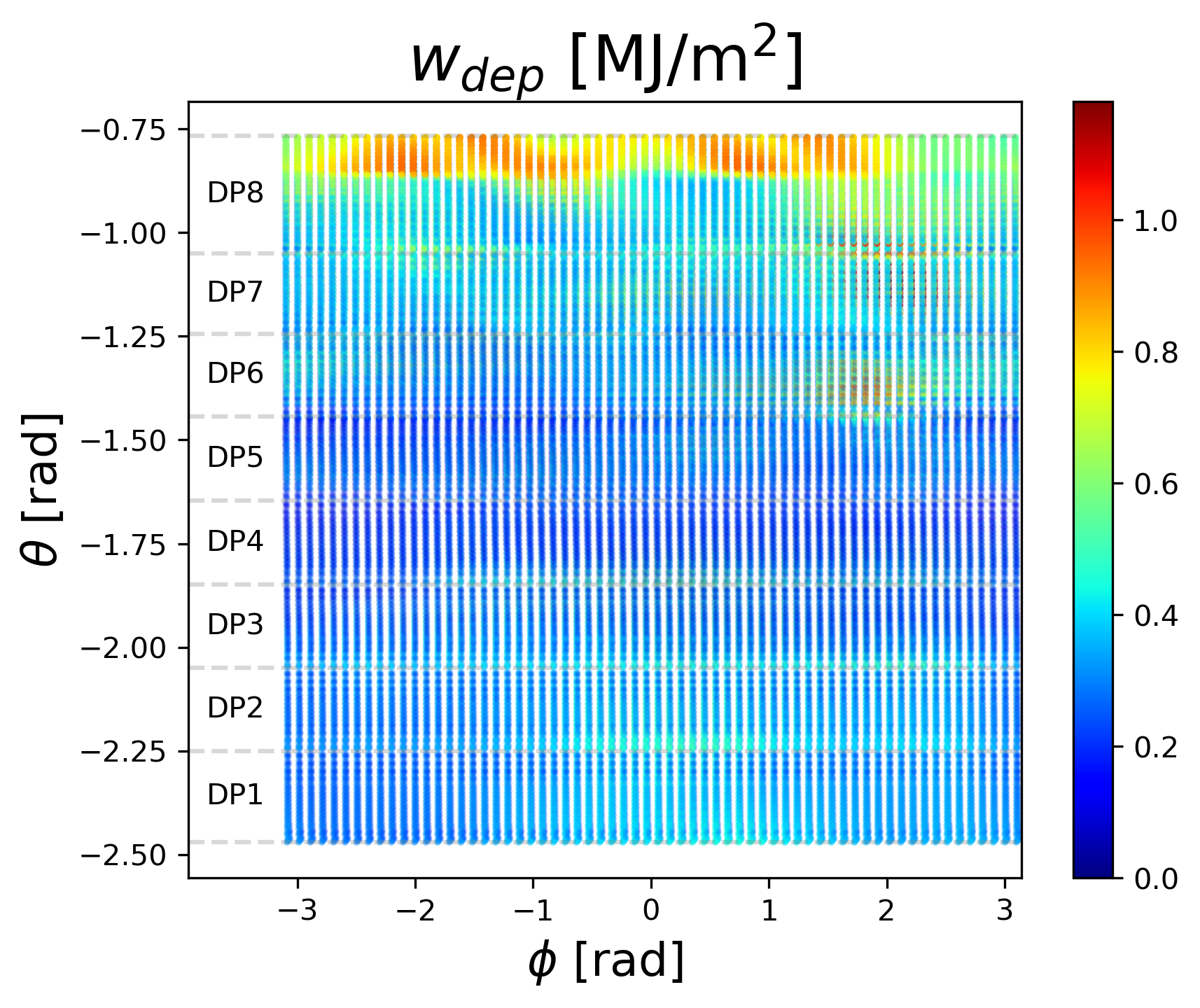} 
\caption{\mod{Time-integrated perpendicular heat flux on the upper DPs for the \#95110 simulation, considering only the TQ (left) and the CQ (right) phases separately.  The plates are displayed in toroidal ($\phi$) and poloidal ($\theta$) angles.} }
\label{fig:wdep_95110_sep}
\end{figure}

\begin{figure}[h]
\centering
\includegraphics[width=0.49\textwidth]{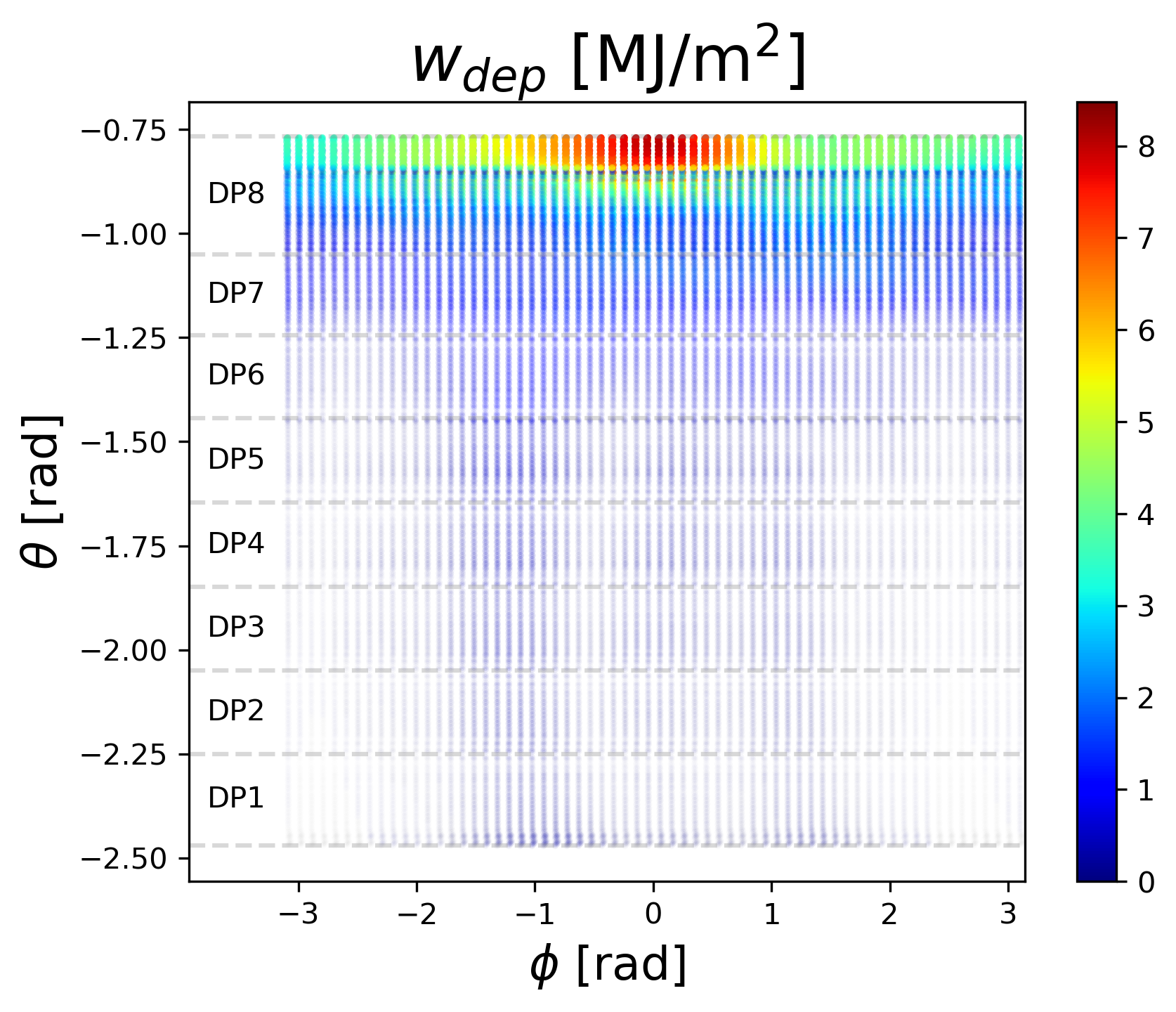} %
\includegraphics[width=0.49\textwidth]{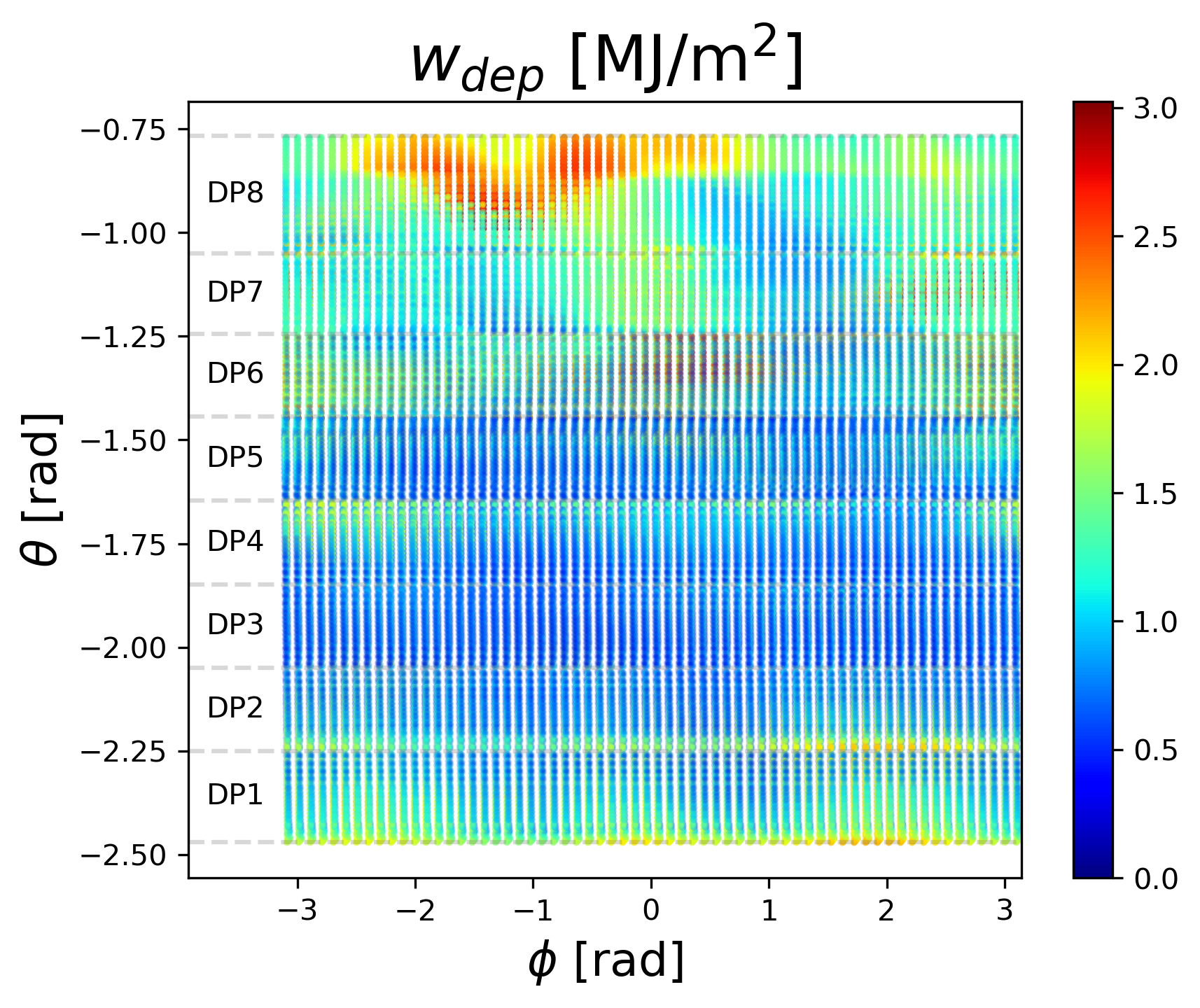} 
\caption{\mod{Time-integrated perpendicular heat flux on the upper DPs for the \#84832 simulation, considering only the TQ (left) and the CQ (right) phases separately.  The plates are displayed in toroidal ($\phi$) and poloidal ($\theta$) angles.} }
\label{fig:wdep_84832_sep}
\end{figure}

\mod{
We now turn to compute an effective energy deposition area, which may provide a practical parametrization for engineering assessments of thermal loads. For this purpose, we time-integrate the perpendicular heat flux on the upper DPs, $q_\perp$, to obtain the deposited energy per unit surface area, $w_{\mathrm{dep}}$. Since the TQ and CQ exhibit markedly different deposition patterns, we evaluate $w_{\mathrm{dep}}$ for each phase separately. The resulting maps are shown in Figs.~\ref{fig:wdep_95110_sep} and \ref{fig:wdep_84832_sep} for pulses \#95110 and \#84832, respectively. We then define an effective deposition area for each phase as
\begin{equation}
A_{\mathrm{eff}} \;=\; \frac{E_{\mathrm{dep}}}{w_{\mathrm{char}}}
\end{equation}
where $E_{\mathrm{dep}}$ is the total energy deposited during the phase (taken here as $E_{th}$ for the TQ and $E_{ohm}$ for the CQ), and $w_{\mathrm{char}}$ is a characteristic surface energy density. We choose $w_{\mathrm{char}} = 0.5\,\max\!\left(w_{\mathrm{dep}}\right)$, which provides a simple and reproducible measure anchored to the peak loading. With this definition, we obtain $A_{\mathrm{eff}}=0.42~\mathrm{m}^2$ and $0.50~\mathrm{m}^2$ for the TQ, and $A_{\mathrm{eff}} =3.9~\mathrm{m}^2$ and $6.2~\mathrm{m}^2$ for the CQ, for \#95110 and \#84832, respectively. Notably, these effective areas depend only weakly on the overall energy content. Note that $E_{th}$ and $E_{ohm}$ differ by roughly factors of 10 and 4 between the two pulses, yet the corresponding $A_{\mathrm{eff}}$ values remain similar.
}

\medskip

\begin{figure}[h]
\centering
\includegraphics[width=0.6\textwidth]{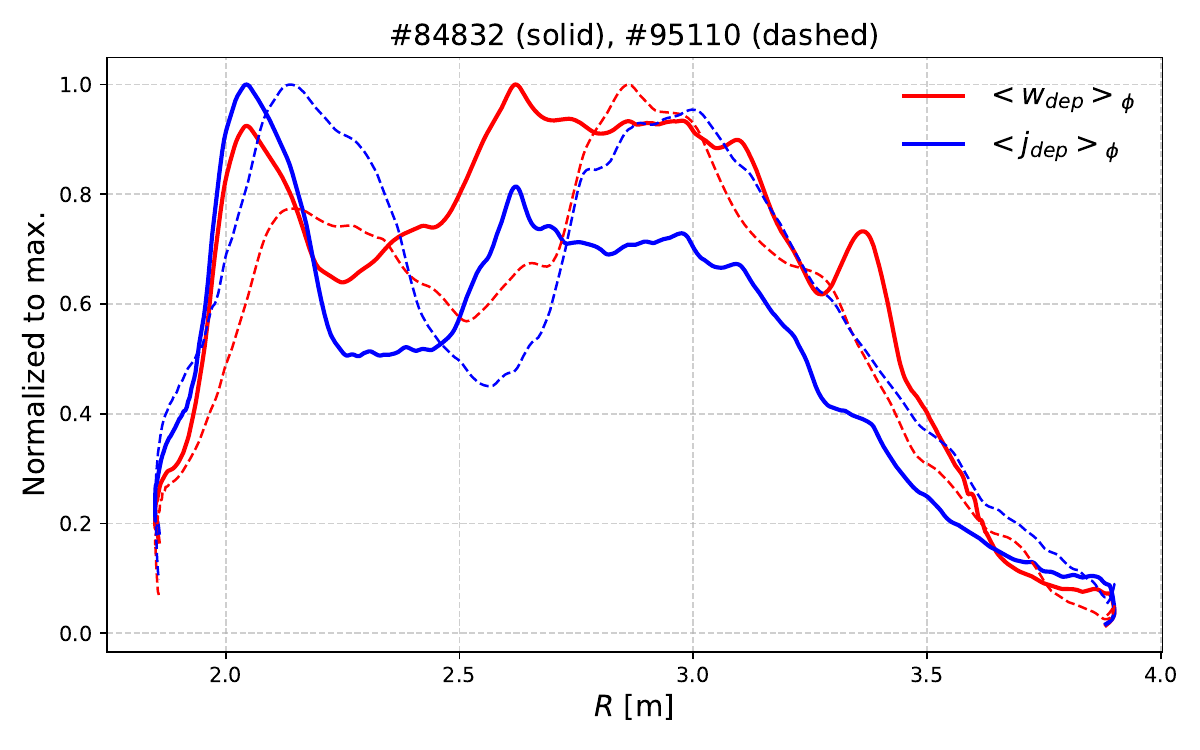}  
\caption{\mod{Time-integrated toroidally averaged profiles on the \textsc{JOREK} boundary for JET pulses \#95110 (dashed) and \#84832 (solid). Shown are the deposited energy density $w_{\mathrm{dep}}=\int q_\perp\,dt$ and the time-integrated normal current density $j_{\mathrm{dep}}=\int J_\perp\,dt$, with $J_\perp=|\mathbf{J}\cdot \mathbf{n}|$, plotted as a function of the major radius $R$. Quantities are normalized to their maximum values.}}
\label{fig:wdep_jdep_JET}
\end{figure}

\mod{
Another practical question is whether the conducted energy and the halo current exhibit similar deposition profiles during the CQ. Consider a given flux tube in the halo region and assume negligible radiation. In this simplified picture, the electron temperature is set by a balance between ohmic heating along the tube and end losses at the wall. The end losses scale as $q_\parallel \propto n_e T_e^{3/2}$ (sheath-limited), while the ohmic heating scales as $\eta J_\parallel^2$ with $\eta\propto T_e^{-3/2}$ (Spitzer resistivity). Combining these scalings yields $q_\parallel\propto\sqrt{n_e}J_\parallel$, suggesting that current and heat should follow comparable spatial footprints. Unfortunately, current measurements are not available at the upper DPs to confirm this finding. We therefore utilize the \textsc{JOREK} simulations as basis, where energy and current profiles have been validated with the available TCs for the energy depostion and MS and RC for the current flow when available. Fig. \ref{fig:wdep_jdep_JET} compares the time-integrated, toroidally averaged profiles of $q_\perp$ and of the normal current density, $J_\perp=|\mathbf{J}\!\cdot\!\mathbf{n}|$, for pulse \#95110 (dashed) and \#84832 (solid). Although the two profiles do not exactly overlap, both quantities exhibit very similar decay lengths. A further notable feature is the width of the profiles, with large radial decay lengths of the order $0.5~\mathrm{m}$.
}

\section{Prediction for a high current ITER VDE}
\label{sec:ITER}

Having established a reasonably successful validation of the modelling workflow against dedicated JET UVDE experiments in Section~\ref{sec:JET}, we now turn to predictions for ITER. We focus on an upward-going UVDE simulated with \textsc{JOREK}, which corresponds to the case with a CQ duration of 240 ms in Ref. \cite{Artola_PPCF_2024}. This case represents the maximum magnetic energy content of ITER disruptions, since it assumes the maximum  $I_p=15$~MA. At the same time, this is an L-mode plasma and the pre-TQ thermal energy is modest, about $25$~MJ, which is much smaller than the $\sim350$~MJ of thermal energy stored in a fully developed $Q=10$ H-mode scenario. Consequently, the results presented here should be regarded as a worst-case scenario for CQ-induced thermal loads, but not for TQ loads. The latter could be considerably more severe in the event of a loss of vertical control in a fully developed $Q=10$ plasma without disruption mitigation, which is a situation that is expected to be extremely rare during ITER operation.

\begin{figure}[h]
\centering
\includegraphics[width=0.49\textwidth]{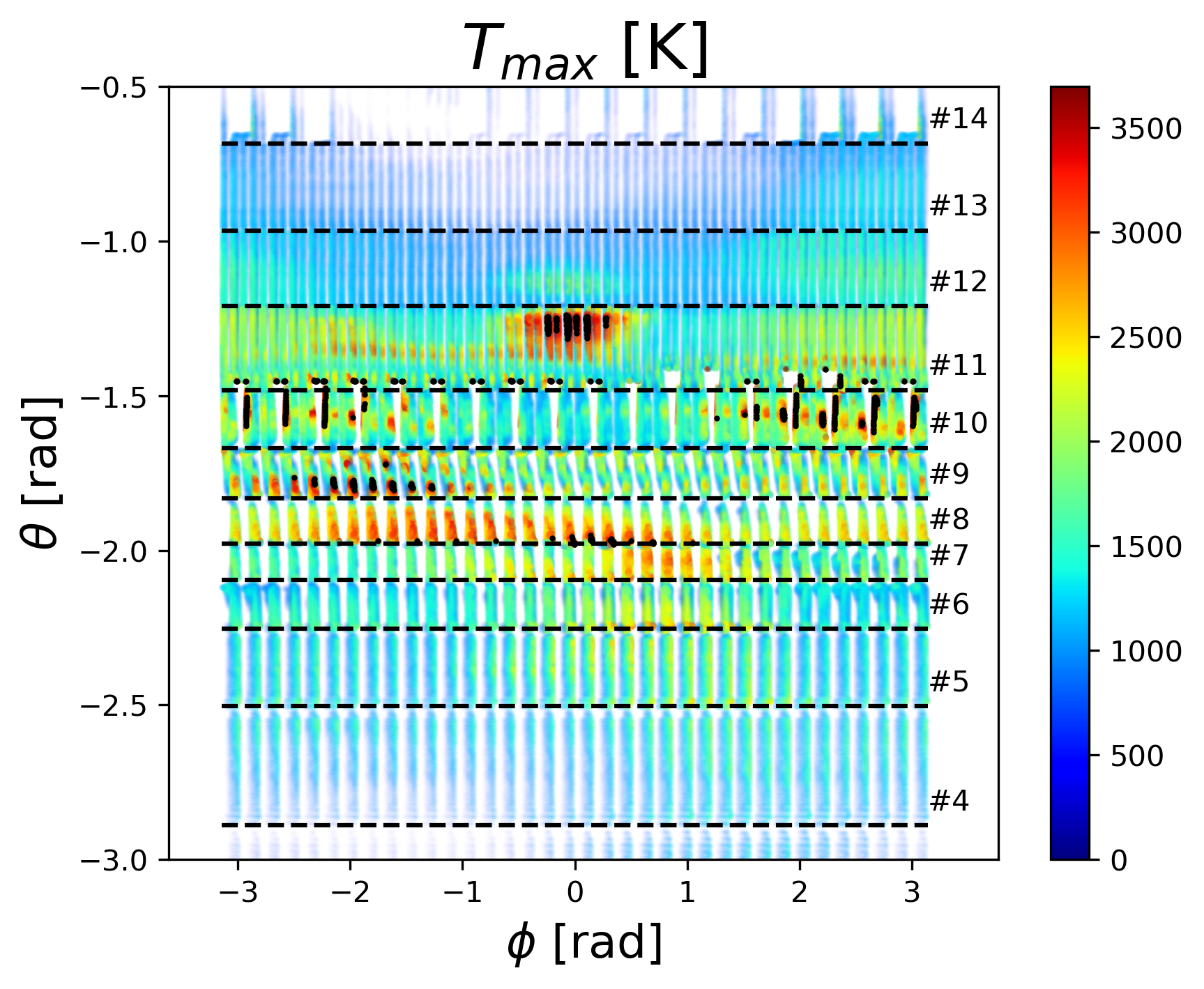} %
\includegraphics[width=0.49\textwidth]{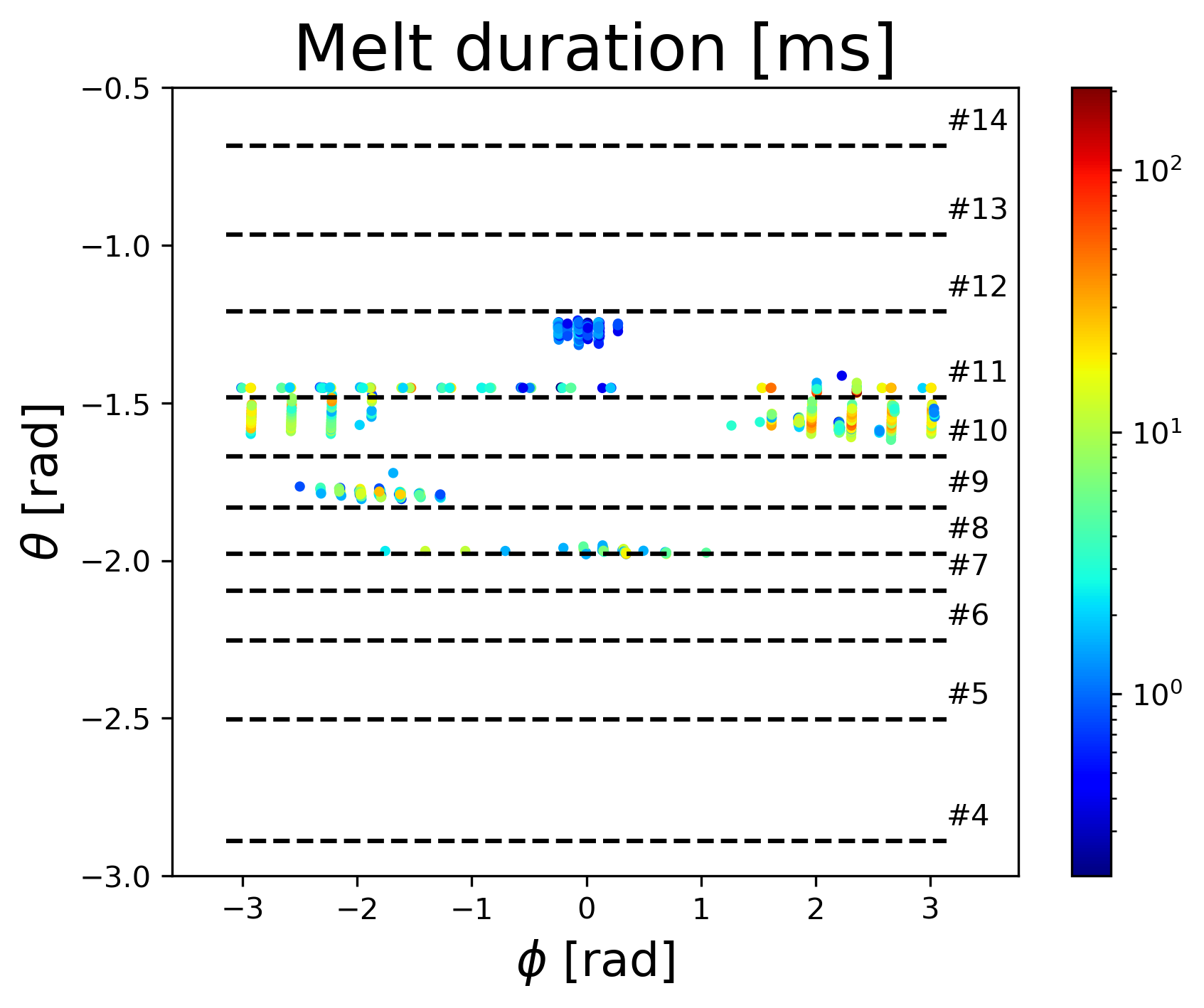} 
\caption{Maximum surface temperature (left) and melt duration (right) on the ITER W FW during a 15~MA upward-going UVDE. The FW surface is represented in toroidal ($\phi$) and poloidal ($\theta$) coordinates, with the poloidal index of the FWPs (\#). Black dots mark elements where the surface temperature exceeds the W melting point. }
\label{fig:Tmax_ITER}
\end{figure}

\medskip

Fig. \ref{fig:Tmax_ITER} (left) presents the maximum surface temperature attained during the disruption, assuming a pre-disruptive wall temperature of $500^{\circ}$C\footnote{This temperature value is expected to be reasonable during burning plasma operation on the upper FWPs heated by stationary heat fluxes on the secondary X-point region (FWPs \#7-9). However, such assumption overestimates the final temperature in other FWPs such as \#10 and \#11.}. The map is subdivided into individual FWPs for reference. The first instances of melting appear during the TQ, which lasts about 6~ms, affecting FWP~\#11. However, in this case the surface remains above the W melt threshold for only $\sim2$~ms, as shown in Fig.~\ref{fig:Tmax_ITER} (right). As the CQ proceeds and the plasma column drifts upward, the load shifts progressively upwards and inwards (with FWP~\#7 being the uppermost FWP receiving significant loads). The most pronounced melting is predicted on FWP~\#10, with melt durations reaching several tens of milliseconds. Importantly, these events are not located on the main areas of the panel surfaces, but rather near the panel edges where the local incidence angle between field lines and the wall normal becomes nearly perpendicular (due to the stronger toroidal shaping in the panel wings). Such conditions occur on FWPs~\#10 and~\#11, where the open geometry around the ITER upper ports allows deeper penetration of magnetic field lines. For the range of FWPs \#8-\#11, localized melting can be observed on the main FWP areas with melt durations typically below 20~ms.

\begin{figure}[h]
\centering
\includegraphics[width=0.49\textwidth]{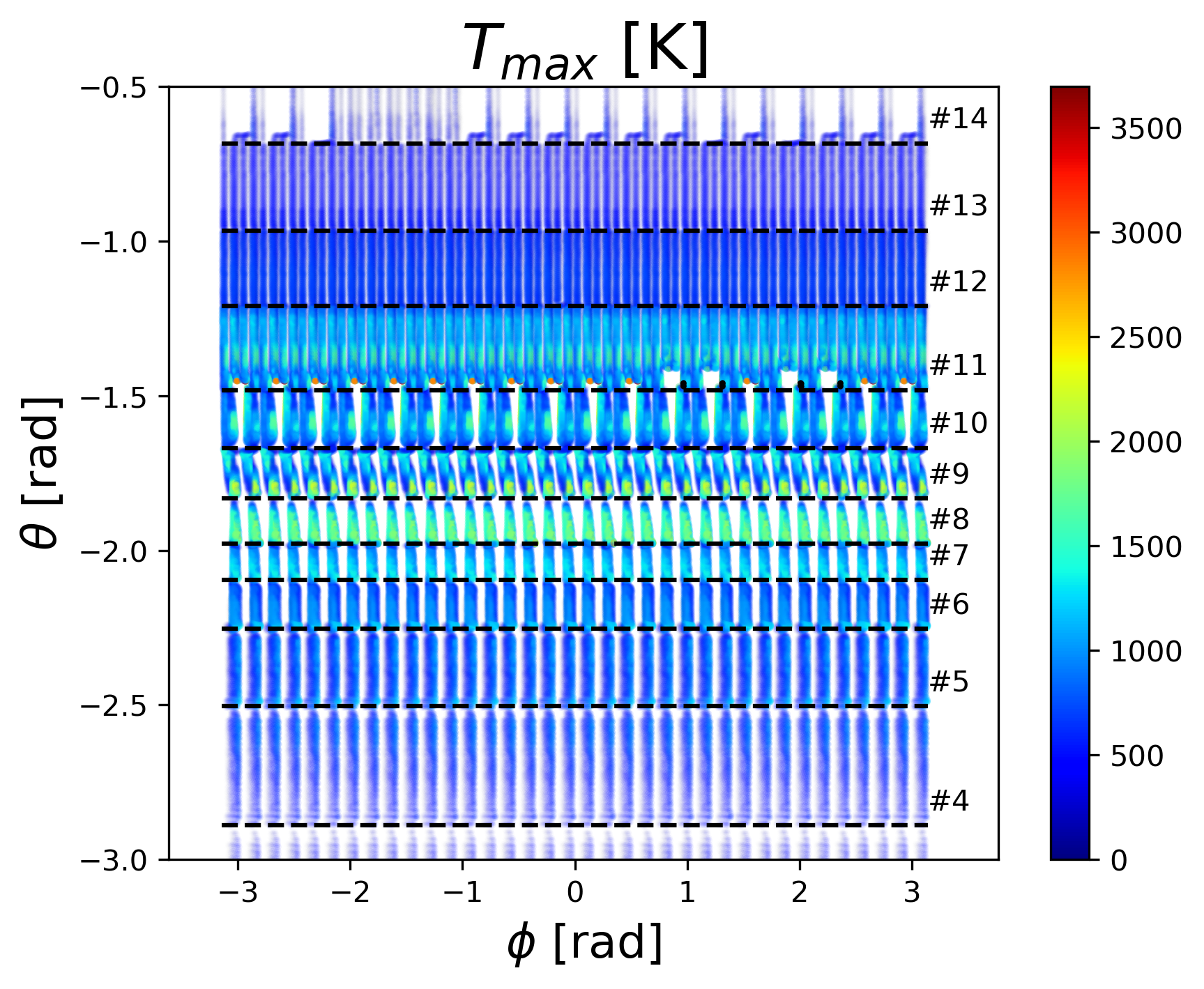} %
\includegraphics[width=0.49\textwidth]{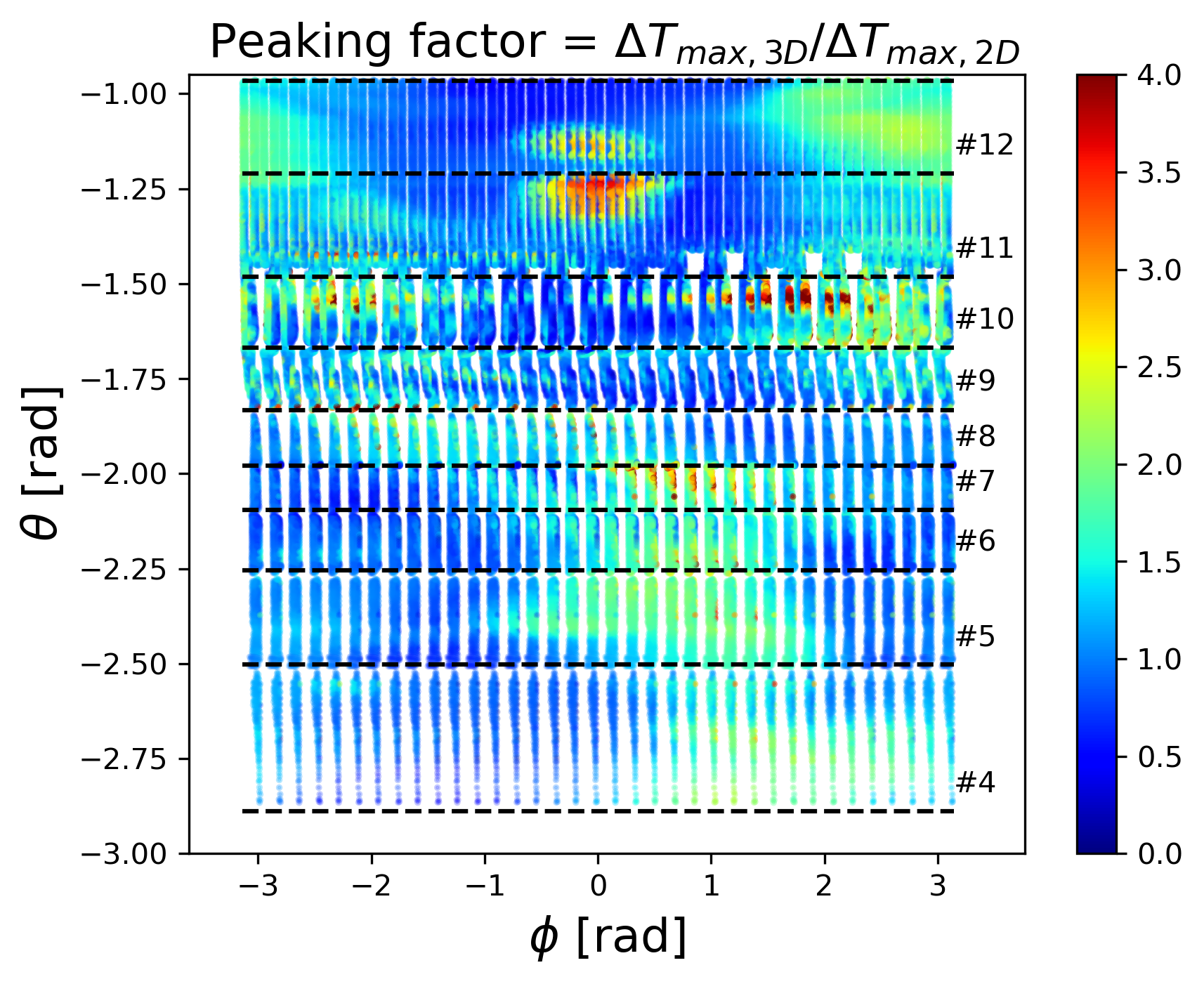} 
\caption{\mod{(Left) Maximum surface temperature for the ITER case in Fig. \ref{fig:Tmax_ITER} when averaging toroidally the magnetic field and $q_\parallel$. The FW surface is represented in toroidal ($\phi$) and poloidal ($\theta$) coordinates, with the poloidal index of the FWPs (\#). Black dots mark elements where the surface temperature exceeds the W melting point. (Right) Toroidal peaking factor of the maximum FW temperature calculated by dividing the temperature increase of the simulation in \ref{fig:Tmax_ITER} by the left panel. }  }
\label{fig:T_2D_peaking}
\end{figure}

\medskip
\mod{
We now turn to assess the impact of toroidal asymmetries on the predicted thermal response. To quantify this effect, we construct a reference ``axisymmetric'' load by toroidally averaging the magnetic field and $q_\parallel$, before performing the field line tracing and thermal calculations. The resulting maximum surface temperature map is shown in Fig.~\ref{fig:T_2D_peaking} (left). In this toroidally averaged case, only a few elements of the edges exceed transiently the W melting temperature, indicating that an axisymmetric model would not predict significant damage for this scenario. We then evaluate the toroidal localization of the loads by comparing the fully 3D simulation to the axisymmetric reference. Specifically, we define a toroidal peaking factor as the ratio of the temperature rise in the 3D case to that obtained with toroidally averaged fields, as shown in Fig.~\ref{fig:T_2D_peaking} (right). The peaking is largest during the TQ, reaching values of $\simeq 3.5$ on panel \#11. During the CQ the asymmetry is weaker but still significant, with typical peaking factors of order 2 on the main impacted panels \#7--\#9. 
}

\begin{figure}[h]
\centering
\includegraphics[width=0.49\textwidth]{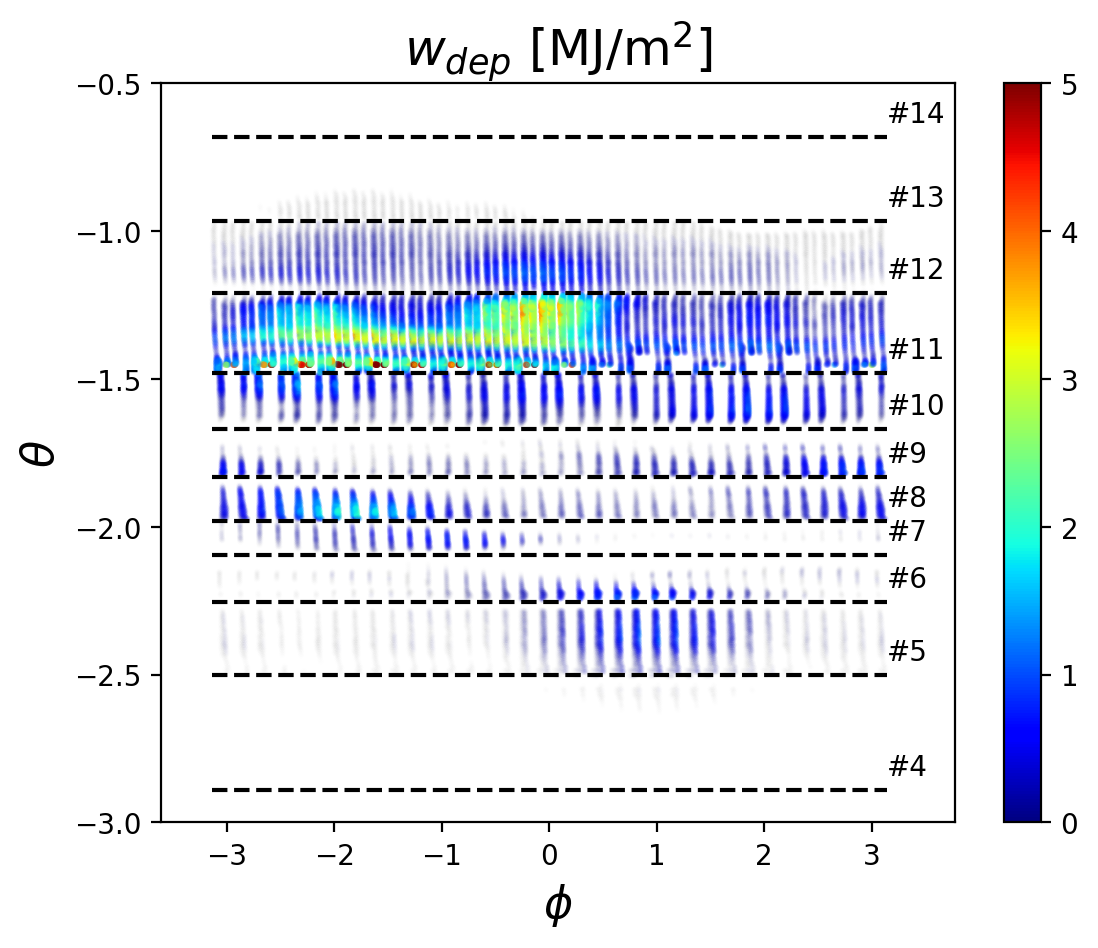} %
\includegraphics[width=0.49\textwidth]{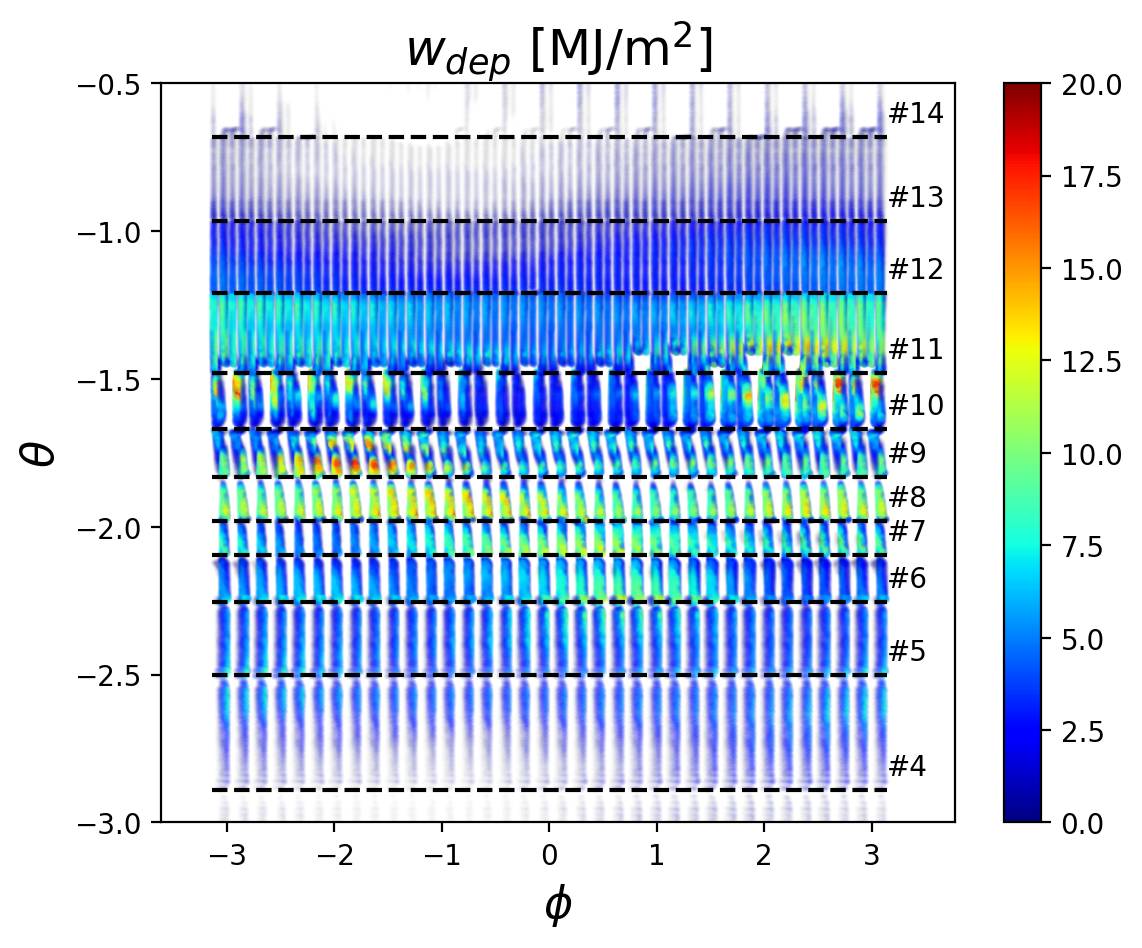} 
\caption{\mod{Time-integrated deposited energy density on the ITER FWPs, computed from $w_{\mathrm{dep}}=\int q_\perp\,dt$ and shown separately for the TQ (left) and CQ (right). The first wall is displayed in toroidal ($\phi$) and poloidal ($\theta$) angles.}}
\label{fig:wdep_ITER}
\end{figure}

\medskip

\mod{
Following the same approach as in Section~\ref{sec:JET_area}, we evaluate the deposited energy density for the ITER case and separate the contributions from the TQ and CQ (Fig.~\ref{fig:wdep_ITER}). The TQ footprint is much more localized, both poloidally and toroidally, than the CQ footprint. However, the absolute loading is substantially larger during the CQ because much more energy is available. Over the main affected panel areas we find $w_{\mathrm{dep}}^{\max}\sim 4~\mathrm{MJ\,m^{-2}}$ during the TQ, compared to $\sim 20~\mathrm{MJ\,m^{-2}}$ during the CQ. This difference is consistent with the available energies, with $25~\mathrm{MJ}$ released during the TQ versus $637~\mathrm{MJ}$ of deposited magnetic energy during the CQ. Using these characteristic values, the corresponding effective deposition areas are $A_{\mathrm{eff}}\approx 12~\mathrm{m}^2$ for the TQ and $A_{\mathrm{eff}}\approx 64~\mathrm{m}^2$ for the CQ.
}

\medskip

\begin{figure}[h]
\centering
\includegraphics[width=0.6\textwidth]{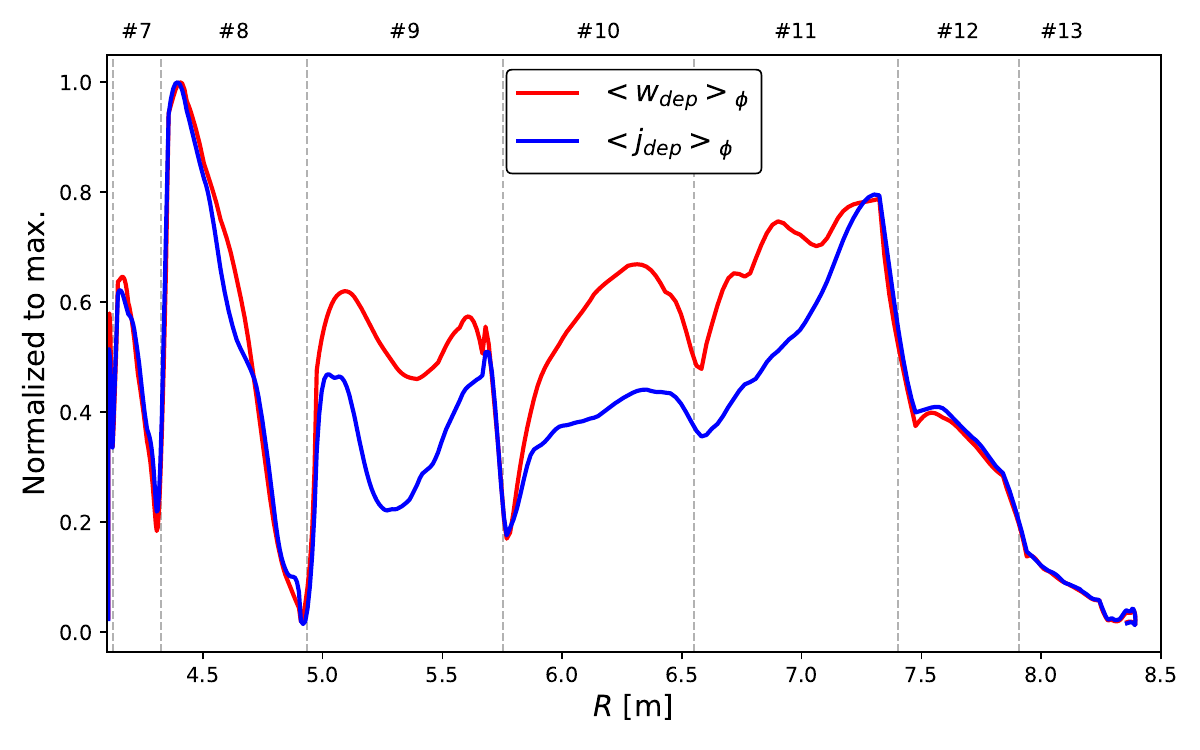}
\caption{\mod{ITER case: time-integrated toroidally averaged profiles on the \textsc{JOREK} boundary. Shown are the deposited energy density $w_{\mathrm{dep}}=\int q_\perp\,dt$ and the time-integrated normal current density $j_{\mathrm{dep}}=\int J_\perp\,dt$, with $J_\perp=|\mathbf{J}\cdot \mathbf{n}|$, plotted as a function of the major radius $R$. Quantities are normalized to their maximum values. The top indices denote the FWP indices for reference.}}
\label{fig:wdep_jdep_ITER}
\end{figure}

\mod{
We now address whether the conducted energy and the halo current exhibit comparable deposition profiles during the CQ. The physical argument and assumptions are identical to those discussed for JET in Section~\ref{sec:JET_area} and are not repeated here.  Fig.  \ref{fig:wdep_jdep_ITER} shows the toroidally averaged, time-integrated profiles of $w_{\mathrm{dep}}=\int q_\perp\,dt$ and $j_{\mathrm{dep}} =\int J_\perp\,dt$ with $J_\perp=|\mathbf{J}\cdot\mathbf{n}|$, extracted along the \textsc{JOREK} domain during the CQ. Consistent with the JET findings, both quantities are expected to display similar spatial footprints and comparable decay lengths, supporting the use of halo current patterns as a proxy for the localization of conducted energy.
}

\medskip

\mod{
One may notice the broadness of the CQ deposition footprint in Fig.~\ref{fig:wdep_ITER}. Since this width is a key input for
engineering assessments, it is important to clarify its physical
origin. A plausible hypothesis is that the footprint is set by the extent to which open field lines penetrate into the confined plasma. In this picture, the broader the region of FW points magnetically connected to the plasma interior, the broader the resulting heat and current deposited on PFCs. We note that a similar physics picture applies to the power deposited in plasma facing components by the application of resonant magnetic perturbations (RMPs) for ELM control (see \cite{Frerichs_PRL_2020} and references therein).  
}

\begin{figure}[h]
\centering
\includegraphics[width=0.32\textwidth]{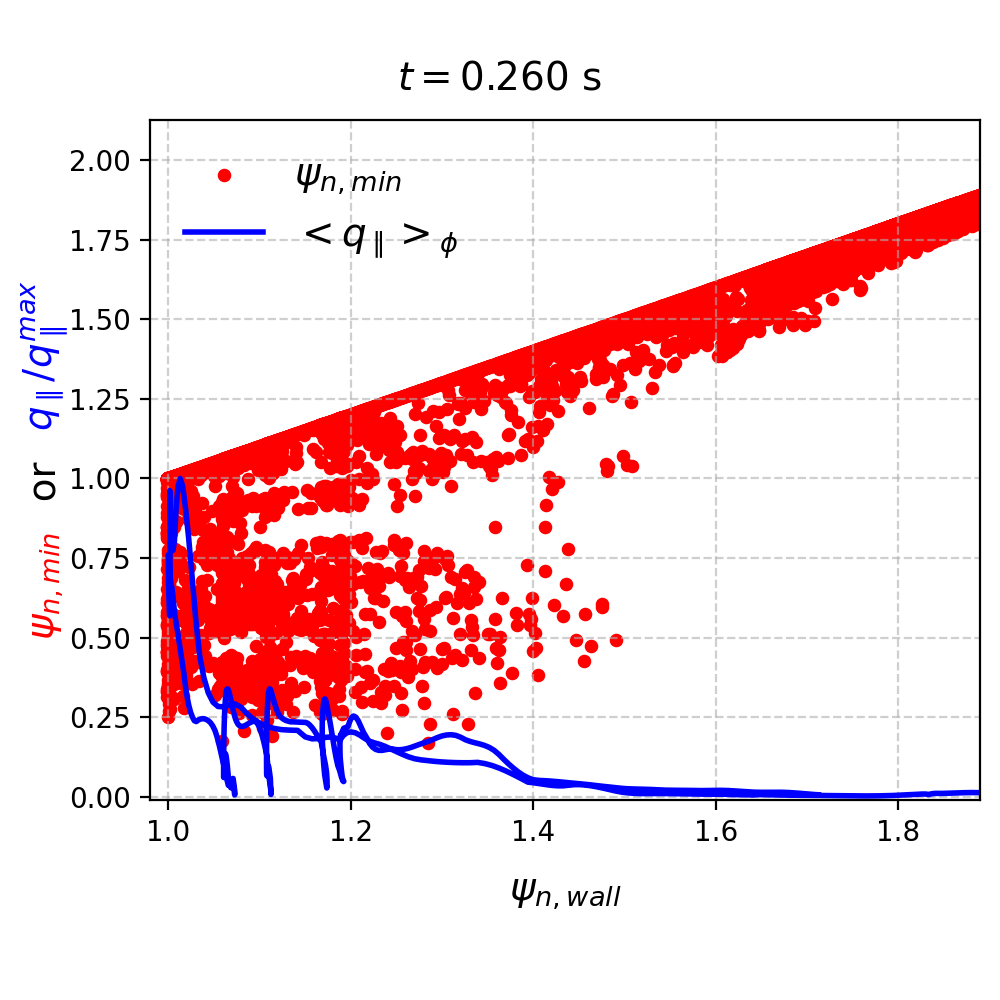} %
\includegraphics[width=0.32\textwidth]{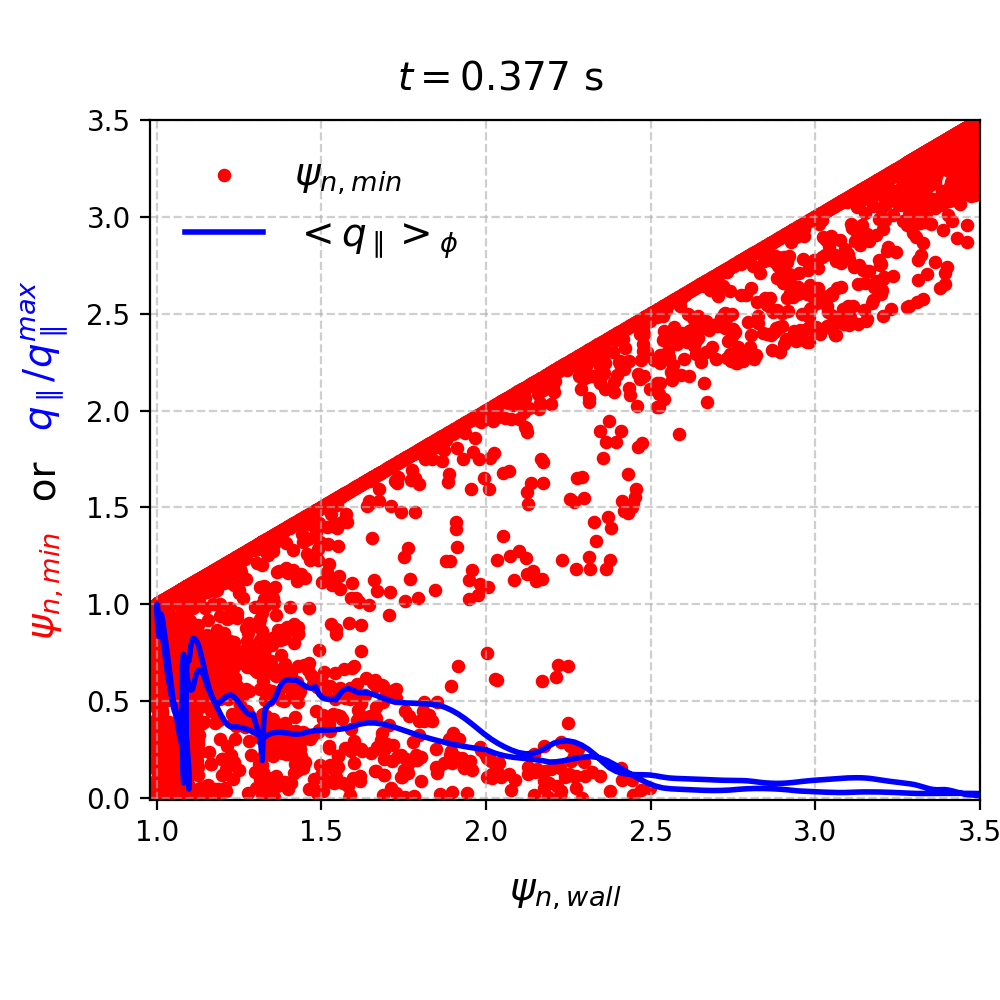} 
\includegraphics[width=0.32\textwidth]{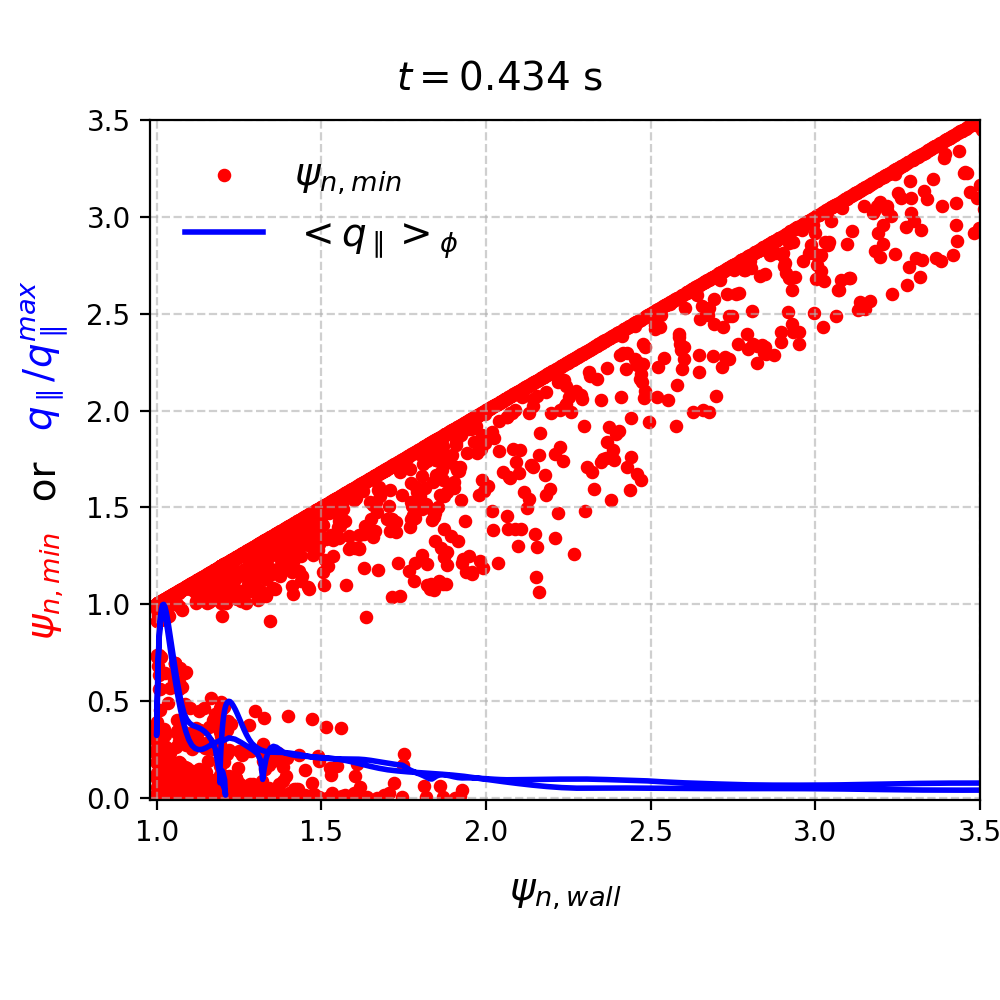} %

\caption{\mod{Relationship between the toroidally averaged $q_\parallel$ at the FW and the field line penetration into the core plasma ($\psi_{n,min}$) for three time points.}}
\label{fig:width_MHD}
\end{figure}

\medskip

\mod{
To test this idea, we trace magnetic field lines from the FW (starting at a given $\psi_{n,\mathrm{wall}}$) towards the plasma and record the deepest penetration they achieve, quantified by the minimum normalized poloidal flux encountered along the trajectory, $\psi_{n,\min}$. For this diagnostic we compute $\psi_n$ using only the axisymmetric component ($n=0$) of the simulation, which provides well-defined flux surfaces that serve as a convenient reference. If the heat flux is primarily controlled by magnetic connection to the core, we expect the largest $q_\parallel$ at wall locations for which the corresponding field lines reach further into the plasma, i.e.\ for which $\psi_{n,\min}$ is significantly below unity. Fig. ~\ref{fig:width_MHD} compares, at three representative times, the toroidally averaged parallel heat flux at the wall, $\langle q_\parallel\rangle_\phi$, with the corresponding $\psi_{n,\min}$ obtained from field line tracing. A clear correlation is observed: the radial extent of the heat flux footprint coincides with the region where field lines satisfy $\psi_{n,\min}(\psi_{n,\mathrm{wall}}) < 1$, i.e.\ where they penetrate inside the last closed flux surface. This indicates that the CQ footprint width in UVDEs is controlled primarily by the 3D MHD-driven deformation of the magnetic topology, which sets how far core-connected field lines spread into the SOL.
}

\medskip

\mod{
An important implication is that purely axisymmetric models cannot
predict this deposition width self-consistently, unless supplemented by
additional information on the amplitude of the non-axisymmetric MHD
perturbations. The strong link between mode amplitude and
effective footprint suggests the development of reduced
descriptions in which physics-informed estimates of the perturbation
strength (given a $q$-profile) are mapped onto an effective connection width (or corresponding transport coefficient) for use in faster engineering-oriented tools such as \textsc{DINA} (or \textsc{JOREK} in its axisymmetric version). In this interpretation, the dominant control of the halo/footprint width is the 3D magnetic connection, rather than solely resistive or diffusive mechanisms previously invoked.}

\section{Conclusions and outlook}
\label{sec:conclusions}

We have combined MHD simulations, field line tracing on a realistic 3D wall, and a transient wall thermal response model to study UVDEs in JET and ITER. Validation against two dedicated JET discharges shows that the simulations reproduce the global evolution of the measured plasma current, current-centroid motion, current asymmetries and yield reasonable agreement with thermocouple-inferred energy deposition. Although boundary conditions are simplified, the CQ appears to be conduction-limited, which reduces sensitivity to such simplifications. Consistent with experiment, melting of Be PFCs is predicted for the high-current JET case and not for the lower-current case; crucially, the melting requires the combined action of the TQ and CQ, with the TQ preheating the surface and the CQ sustaining and amplifying the temperature rise. These validated heat-load cases may provide physics-based inputs for dedicated melt-motion modelling (e.g. \textsc{MEMENTO} \cite{Paschalidis_FED_2024}).

\medskip
Applying the workflow to ITER, we considered a 15 MA unmitigated upward VDE with a 240 ms CQ. The main chamber W FWPs experience only marginal melting with short durations, whereas more substantial melting (which can reach melt durations of several tens of milliseconds) is localized to exposed panel edges near the upper ports where field lines intersect the wall at near-normal incidence. Toroidal plasma asymmetries amplify local peak surface temperatures relative to axisymmetric assumptions by up to a factor of 2 during the CQ and 3.5 during the TQ, highlighting the need for fully 3D treatments. At the same time, the MHD dynamics broaden the heat flux profiles compared with axisymmetric transport/equilibrium workflows (e.g. \textsc{DINA} and \textsc{TOKES}), which would otherwise over-localize the loads. For example, 2D \textsc{TOKES} simulations for the W FWPs predicted CQ melting already at 10 MA considering a fixed equilibrium and poloidal energy deposition width $\lambda_E=3.5$ cm. Despite a two-fold increase in energy (15 MA) and toroidal localization, the \textsc{JOREK} simulations show marginal melting due to the moving equilibrium and MHD spreading, leading to much broader $\lambda_E$ in the range of several tens of cm. 

\medskip

\mod{
For engineering-oriented metrics, we quantified effective deposition areas for ITER from the deposited energy density maps. Using the characteristic surface loadings obtained in Fig.~\ref{fig:wdep_ITER}, we find $A_{\mathrm{eff}}\approx 12~\mathrm{m}^2$ for the TQ and $A_{\mathrm{eff}}\approx 64~\mathrm{m}^2$ for the CQ. In addition, the simulations show that heat and current channels are tightly linked during the CQ: the time-integrated profiles of deposited energy density and normal current density exhibit similar spatial footprints and decay lengths (Fig.~\ref{fig:wdep_jdep_ITER}), supporting the use of halo current patterns as a proxy for the localization of conducted energy when direct heat flux measurements are not available. We also provided evidence that the broadening of the CQ footprint is set by magnetic connection rather than solely by resistive or diffusive broadening. Field line tracing demonstrates that the radial extent of $\langle q_\parallel\rangle_\phi$ coincides with the region where field lines intersecting the wall penetrate inside the last closed flux surface, $\psi_{n,\min}<1$ (Fig.~\ref{fig:width_MHD}). This links the effective footprint width directly to the extent over which field lines open and connect to the confined plasma, i.e.\ to the 3D MHD-driven deformation of the magnetic topology.
}

\medskip

\mod{
Overall, this state-of-the-art workflow highlights that the ITER W FW is markedly more resilient to disruption heat loads than the previous Be armoured wall, easing CQ thermal-mitigation requirements. Because radiation and impurity effects are neglected here, the reported temperatures and melt durations may also be regarded as conservative.} Future work should extend the physics fidelity of the \textsc{JOREK} simulations by including radiation and impurities, refining boundary conditions (sheath physics and density evolution), and exploring TQ-dominated scenarios with higher thermal energies. Coupling the present heat-load predictions to melt-evolution solvers will enable assessments of melt-layer motion, droplet ejection, and lifetime impacts. \textsc{TOKES} simulations of the TQ with input guided by these results, have been produced and will be reported in a forthcoming publication. The impact of W evaporation and  plasma radiation as a self-mitigation mechanism (vapour shielding) can also be investigated with \textsc{TOKES}.

%

\ack{ITER is the Nuclear Facility INB No. 174. This work explores the physics processes during plasma operation of the tokamak when disruptions take place; nevertheless the nuclear operator is not constrained by the results presented here. The views and opinions expressed herein do not necessarily reflect those of the ITER Organization. The simulations presented here have been performed using the JFRS-1 and the ITER SDCC clusters. This work has been carried out within the framework of the EUROfusion Consortium, funded by the European Union via the Euratom Research and Training Programme  (Grant Agreement No 101052200 — EUROfusion). Views and opinions expressed  are however those of the author(s) only and do not necessarily reflect those of the  European Union or the European Commission. Neither the European Union nor the  European Commission can be held responsible for them.}


%

\bibliographystyle{unsrt}
\bibliography{biblio.bib}{}

\end{document}